\documentclass[a4paper, amsfonts, amssymb, amsmath, reprint, showkeys, nofootinbib, twoside, floatfix]{revtex4-1}
\usepackage{xcolor}
\usepackage{physics}
\usepackage[english]{babel}
\usepackage[utf8]{inputenc}
\usepackage[colorinlistoftodos, color=green!40, prependcaption]{todonotes}
\usepackage[pdftex, pdftitle={Article}, pdfauthor={Author}]{hyperref} 
\def \reals{{\mathbb R}}

\def \EvalM{\biggr\rvert_M}
\def \EvalQ{\biggr\rvert_Q}
\def \evalM{|_M}

\newcommand{\td}[2]{\frac{d #1}{d #2}}

\newcommand{\pd}[2]{\frac{\partial #1}{\partial #2}}
\newcommand {\pdd}[2]{\frac{\partial^{2} #1}{\partial #2^{2}}}
\newcommand {\pdm}[3]{\frac{\partial^{2} #1}{\partial #2\partial #3}} 

\newcommand {\intz}[1]{\int_{0}^{Lz}#1 dz} 

\def \hetaO{\hat{\eta}_0}   
\def \heta{\hat{\eta}}

\def \hphi{\hat{\varphi}}

\def \Lop{L}     
\def \Lopc{\mathcal{L}}    

\begin{document}
\title{Following marginal stability manifolds in quasilinear dynamical reductions of multiscale flows in two space dimensions}

\author{
  Alessia Ferraro $^{1,2}$,
  Gregory P. Chini $^3$, 
  T. M. Schneider $^1$, 
  }

\affiliation{
$^1$Emergent Complexity in Physical Systems Laboratory (ECPS), 
\'Ecole Polytechnique F\'ed\'erale de Lausanne, CH-1015 Lausanne, Switzerland\\
$^2$Nordita, Royal Institute of Technology and Stockholm University, Stockholm 106 91, Sweden\\ 
$^3$Department of Mechanical Engineering and Program in Integrated Applied Mathematics, University of New Hampshire, Durham,
NH 03824, USA\\
}

\date{\today} 

\begin{abstract}
A two-dimensional (2D) extension of a recently developed formalism for slow--fast quasilinear (QL) systems subject to fast instabilities is derived. 
Prior work has demonstrated that the emergent dynamics of these systems is characterized by a slow evolution of (suitably defined) mean fields coupled to marginally stable, fast fluctuation fields. 
By exploiting this emergent behavior,
an efficient hybrid fast-eigenvalue/slow-initial-value solution algorithm can be developed in which the amplitude of the fast fluctuations is slaved to the slowly evolving mean fields to ensure marginal stability---and temporal scale separation---is maintained. For 2D systems that are spatially-extended in one direction, the fluctuation eigenfunctions are labeled by their Fourier wavenumbers characterizing spatial variability in that direction, and the marginal mode(s) also must coincide with the fastest-growing mode(s) over all admissible Fourier wavenumbers.
Here, we introduce two alternative but equivalent procedures for deriving an ordinary differential equation governing the slow evolution of the wavenumber of the fastest-growing fluctuation mode that simultaneously must be slaved to the mean dynamics to ensure the mode has zero growth rate. We illustrate the procedure in the context of a 2D model partial differential equation that shares certain attributes with the equations governing strongly stratified shear flows and other strongly constrained forms of geophysical turbulence in extreme parameter regimes. 
The slaved evolution follows one or more marginal stability manifolds, which constitute select state-space structures that are not invariant under the full flow dynamics yet capture quasi-coherent structures in physical space in a manner analogous to invariant solutions identified in, e.g., transitionally-turbulent shear flows. Accordingly, we propose that marginal stability manifolds are central organizing structures in a dynamical systems description of certain classes of multiscale flows in which scale separation justifies a QL approximation of the dynamics. 

\end{abstract}


\maketitle

\section{Introduction}
\label{sec:intro}

Forced--dissipative spatio-temporally chaotic dynamical systems, such as turbulent fluid flows, are often characterized by the spontaneous emergence of self-organized quasi-coherent patterns.
These patterns are particularly evident in \emph{anisotropic} turbulent flows ranging from wall-bounded shear flows in the transitionally turbulent regime \cite{Reynolds1883, Avila2023, Emmons1951, Coles1966, Wu2023, Manneville2001,Prigent2003,Avila2011,Tuckerman2020,Duguet2010} to naturally-occurring turbulent flows in the Earth's oceans and atmosphere \cite{McWilliams1985, McWilliams1994} and in stellar interiors \cite{Spiegel1992, Garaud2018, Alisse2000}.
In the latter case, the presence of a strong physical constraint, e.g., strong stratification, strong rotation and/or a strong magnetic field, restrains the turbulent motion, often inducing a spatial and temporal scale separation in the dynamics \cite{Julien2007}. A paradigmatic example can be found in oceanic and atmospheric turbulence, where the presence of strong stable density stratification gives rise to large-scale anisotropic structures that are coupled to small-scale instabilities, with dynamics occurring on disparate scales (e.g., tens to hundreds of kilometers and tens of meters, respectively, in the Earth's oceans) \cite{Billant2001, Gregg1987, Fitzgerald2018, Fitzgerald2019}.
The mechanisms driving the spontaneous emergence of recognizable patterns in spatio-temporally chaotic or turbulent systems is of fundamental interest as a problem in pattern formation. Moreover, the resulting quasi-coherent dynamical structures also significantly impact the global transport properties of the system \cite{Ivey2008, Ferrari2009, Spiegel1992}. Consequently, quantitatively characterizing coherent structures and understanding their formation and self-sustenance is a crucial aspect of modeling anisotropic multiscale flows. 

For flow parameters for which direct numerical simulations (DNS) of the governing partial differential equations (PDEs) are feasible, these simulations can reveal the signatures of self-organized structures and enable their statistical characterization. Numerical simulation alone, however, almost never provides direct access to individual coherent structures or reveals the mechanisms underlying their emergence and dynamics. Consequently, a variety of complementary mathematical approaches has been developed and employed to better understand quasi-coherent spatio-temporal structures. Among these approaches are linear methods, including resolvent analysis, dynamic mode decomposition, secondary stability theory, and transient growth analysis \cite{McKeon2017, Sharma2013, Moarref2014, McKeon2023, Rowley2017, Rowley2009, Schmid2007, Schmid2022}.

In addition, fully nonlinear methods from dynamical systems theory have been used to \emph{directly} access coherent spatio-temporal structures \cite{Kawahara2012, Graham2021, Crowley2022, chaosbook}. This approach, which to date has been applied primarily to spatio-temporally chaotic shear flows in the transitional regime, is based on the identification of fully nonlinear (two- or three-dimensional) equilibria, traveling waves, periodic orbits, and other non-chaotic time-invariant solutions of the governing equations. These invariant solutions are unstable yet, together with their their entangled stable and unstable manifolds, they organize the state space dynamics so that turbulence can be viewed as a chaotic walk among invariant states. Invariant solutions thus provide a direct link between a precisely defined solution of the fully nonlinear evolution equations and an observed coherent structure in the flow \cite{Nagata1990, Kerswell2005, Eckhardt2007, Cvitanovic2013, Chandler2013, Kawahara2012}. To highlight this link, invariant solutions---in particular, equilibria and travelling waves---have also often been termed `exact coherent structures' or `exact coherent states' (ECS) \cite{Waleffe1998, Waleffe2001, Graham2021}. 
Notably, for wall-bounded shear flows transitioning to turbulence, the dynamical systems approach based on invariant solutions has helped to rationalize many observed coherent patterns. 
The growing set of identified invariant solutions includes, for example, those capturing streaks and vortices in minimal flow units of parallel shear flows \cite{Nagata1990, Kawahara2001, Waleffe1998, Waleffe2001, Waleffe2003, Gibson2008, Park2015}, and those underlying laminar-turbulent patterns in transitional Couette flow \cite{Schneider2010, Gibson2016, Gibson2014}, including solutions representing oblique stripes in both Couette \cite{Reetz2019a, Reetz2020c} and channel flow \cite{Paranjape2020}. Similarly, traveling waves and relative periodic orbits of pipe flow have been identified and capture the inner structure and localization of puffs \cite{Faisst2003, Wedin2004, Hof2004, Avila2013, Budanur2017a}. Recently, invariant solutions underlying many patterns in convectively driven flows have been characterized \cite{Reetz2020, Reetz2020e}, and there is growing evidence that both large and small-scale structures in parallel boundary layers may be described by traveling wave solutions \cite{Yang2019, Azimi2019}. 
    
Despite this success, a link between self-organized coherent structures and invariant solutions is not readily established for systems exhibiting strong spatial and temporal scale separation in extreme parameter regimes, including for a variety of forms of geophysical turbulence. Indeed, the existence of structures at various scales within these systems, coupled via interactions mediated by mechanisms such as ``Reynolds stress'' (or other flux) divergences capturing the average effect of one scale on another, renders the pursuit of invariant solutions challenging. ECS are, by definition, invariant under the dynamics across \emph{all} scales. 
Consequently, it is not clear that this classical notion of invariant solutions is a useful construct for describing the coherent structures in systems with strong scale separation, where recognizable structures emerge on one scale but are advected, modulated, and often eventually destroyed by the dynamics on other scales. 

Accordingly, we aim to identify a different class of structures within the system state-space that, unlike invariant solutions, capture multiscale, and often intermittent, coherent patterns. These state-space structures will enable us to generalize the traditional dynamical systems approach based on invariant solutions to chaotic systems with scale separation within the appropriate mathematical context to describe these systems: namely, quasilinear reductions that exploit asymptotic separation of spatial and temporal scales. Specifically, we identify \emph{marginal stability manifolds} that are \emph{not} invariant under the dynamics but constitute subsets of state-space that guide the evolving trajectory and along which the dynamics simplifies. More specifically, the evolution on one scale (typically associated with fast and small-scale `fluctuations') becomes slaved to and thus parametrically dependent upon a slowly evolving, coarse-grained mean field.

In the context of constrained fluid turbulence, a prevalent feature of many geophysical and astrophysical systems, the emergent scale separation can be exploited using asymptotic techniques to systematically derive reduced PDE models. The reduced PDEs are analytically and numerically easier to treat than the original flow equations, yet still capture key features of the full dynamics \cite{Julien2007,Michel2019,Chini2014b,Chini2022}. Many such asymptotic reductions are of slow--fast quasilinear (QL) form: having decomposed all fields into a slowly evolving mean and fast fluctuations, the fluctuation/fluctuation nonlinearities are found to be negligible in the strong-constraint limit \emph{except} where those interactions feed back upon the slowly-evolving mean fields. In contrast to QL or `mean-field' reductions \cite{Herring1963} often invoked as useful but ultimately \emph{ad hoc} approximations, here the scale-separated QL structure naturally emerges from a formal multiple scales analysis of the primitive equations, and is therefore asymptotically justified.

Within a scale-separated QL formulation in which the fast dynamics is not directly externally forced, the homogeneous linear equation governing the fluctuation dynamics does not constrain the fluctuation amplitude. To close the problem and fix the amplitude, \citet{Michel2019} proposed a condition that directly follows from the asymptotic temporal scale separation: namely, the propagation of marginal stability 
of the fluctuation dynamics
that ensures the preservation of the postulated scale separation. These authors further demonstrate that a hybrid fast-eigenvalue/slow-initial-value algorithm can be developed for evolving the QL system in time. In this approach, the initial-value problem for the fast fluctuations is replaced with an eigenvalue problem in which the mean fields are locally frozen on the fast time scale characterizing the fluctuation dynamics. Thus, the linearized evolution of the fast dynamics is described by modal solutions that, in principle, can exponentially grow or decay on the fast temporal scale. In turn, the mean fields are evolved on a slow time scale using the leading eigenfunctions of the linear operator controlling the fluctuation dynamics to compute the spatial structure of the fluctuation/fluctuation nonlinear feedback on the mean.

While the fast exponential decay of the fluctuations yields no appreciable (sustained) feedback on the mean dynamics, exponential growth of the fluctuations would compromise the convergence of this coupling term; in practice, the purportedly `slow' variable would be forced to respond on the fast temporal scale. The first scenario results in an uncoupled system in which the mean field evolves without any feedback from the fluctuations; the second invalidates the posited scale separation underlying the systematic QL reduction. It follows that the coupling between the fluctuations and the mean fields is only possible when the fast dynamics is constrained to evolve along a marginal stability manifold. This constraint can be leveraged to evaluate the otherwise indeterminate fluctuation amplitude (again noting that the eigenvalue problem for the fluctuation fields is linear and homogeneous). As shown by \citet{Michel2019}, due to the potential exponential growth of the fast fluctuations, the `usual' approach for determining the slow evolution of the fluctuation amplitude---i.e., proceeding to higher-order in the asymptotic analysis of the fluctuation system and ensuring a solvability condition is satisfied to prevent the secular growth of higher-order fluctuation fields \cite{Bender1999, Hinch1991}---is untenable for slow--fast QL systems subject to fast instabilities. Instead, \citet{Michel2019} propose 
a new integration strategy that determines the slow evolution of the fluctuation amplitude by ensuring that whenever the leading fast mode becomes marginally stable its amplitude is slaved to the mean dynamics to maintain a zero growth rate in fast time. These authors demonstrated the efficacy of their multi-time-scale algorithm for a simple one-dimensional model problem. A primary virtue of their formulation is that the fast dynamics need not be temporally resolved.

More recently, \citet{Chini2022} also successfully applied the slaving approach to an idealized model of a physical system: body-forced, two-dimensional (2D) stably stratified flow. This demonstration confirms the novel methodology's potential utility for the investigation of a range of constrained geophysical flows. In addition to enabling the study of dynamics in extreme parameter regimes that remain computationally inaccessible via DNS, marginal stability manifolds facilitate characterization of the dynamically relevant state-space structures arising in complex systems exhibiting strong scale separation. Since the flow complexity collapses when the fluctuations 
become slaved to the slowly evolving mean, evolution along the marginal stability manifold may be expected to correspond physically to the emergence of quasi-coherent flow structures. Consequently, we propose that marginal stability manifolds are the appropriate state-space structures to capture coherent patterns in multiscale flow problems described by slow--fast QL reductions of the governing PDEs.

For 2D systems that are spatially-extended in one `horizontal' direction, an additional challenge arises: the fast fluctuation fields also must be locally marginally stable over Fourier wavenumbers $k$ characterizing spatial variability in that direction. 
This requirement introduces a computationally expensive step into the hybrid eigenvalue/initial-value algorithm, as the wavenumber(s) of the mode(s) that become unstable must first be identified before  appropriately slaving the mode(s) to the mean fields. The marginal mode(s) must not only have zero growth rate(s) but must also be the fastest growing mode(s) over all admissible wavenumbers $k$ (i.e., must be a global maximum of the dispersion relation $\Re{\sigma(T,k)}$, where $\sigma$ is the complex growth rate and $T$ is the slow time variable). In \citet{Chini2022}, this task was accomplished in brute-force fashion by solving multiple eigenvalue problems at each slow time instant to locate the fastest-growing mode and then slaving that mode to the mean fields to ensure its growth rate vanished. Here, we will demonstrate that a slow evolution equation for the wavenumber of the fastest-growing, marginal mode can be self-consistently derived, thereby obviating the need to repeatedly solve eigenvalue problems for a potentially large number of $k$ values. Indeed, a chief virtue of the slow--fast QL reduction is that, in principle, $k$ need \emph{not} be quantized; that is, dynamics and pattern formation in arbitrarily (horizontally) extended domains can be simulated, unlike DNS performed in periodic domains of fixed size. In this case, the brute-force search procedure used in \citet{Chini2022} is particularly expensive, highlighting a key advantage of the algorithm developed in this work. To develop an efficient algorithm enabling the prediction of the wavenumber of the fastest-growing, marginal mode, we consider a 2D model PDE specifically designed to capture key features of the stratified flow problem while limiting the overall complexity of the analysis. 

The remainder of the manuscript is organized as follows. We begin by introducing the fully nonlinear 2D PDEs governing a model system and their reduced counterparts, obtained via multiscale analysis. Subsequently, we revisit the essential steps of the QL procedure for the slaving of fluctuations, as first described in \citet{Michel2019}. We emphasize the conceptual distinctions from the one-dimensional case and present an effective methodology for predicting the wavenumber of the fastest growing mode in both stable and marginally-stable scenarios. Notably, in conditions of marginal stability, we illustrate how the evolution equation for the wavenumber of the marginal mode can be derived using two mathematically distinct yet equivalent approaches, confirming the accuracy of the outcome. We then present numerical results of the extended QL procedure together with a validation study using DNS of the fully nonlinear system. Finally, we provide an interpretation of the marginal stability manifold approach, enabled by the algorithmic advances developed herein, in terms of a dynamical systems description of chaotic flows in the presence of strong scale separation. We conclude with a discussion of  future research efforts that are necessary for the proposed methodology to be profitably applied to constrained geophysical turbulence in physically realistic parameter regimes. 

\noindent
\section{Model equations and QL methodology}
\label{sec_equations}

The two-dimensional model problem under consideration couples the evolution of a slowly-evolving field $U(x,z,t)$ with the fast dynamics of a fluctuation field $\eta(x,z,t)$, as governed by the following dimensionless PDEs:

\begin{equation}
\label{Nonlin_U}
   \pd{U}{t} + U\pd{U}{x} = F(x,z,t)-\nu U-\varepsilon^2\Big(\pd{\eta}{x}\Big)^2 + D\nabla^2 U, 
\end{equation}
\begin{equation}
\label{Nonlin_eta}
    \varepsilon\pd{\eta}{t} = -\eta -\varepsilon^2U\pdd{\eta}{x} - \varepsilon^4\dfrac{\partial^4\eta}{\partial x^4} + \pdd{\eta}{z}-\varepsilon\eta^3 \text{,}
\end{equation}
which are posed on a periodic domain $[0,L_x)\times[0,L_z)$. With shear flows in mind, $x$ will be referred to as a `horizontal' coordinate, while $z$ is a coordinate in which the slow field $U$ is `sheared'. The time variable is $t$. $F(x,z,t)$ is a prescribed, time-dependent external forcing, while $\nu$ is a Rayleigh (viscous) damping coefficient and $D$ a diffusion coefficient. The scale separation between the two dynamics is effected via the introduction of the small parameter $\varepsilon$.\\

The evolution equation for the slow field $U$ has certain evident commonalities with the (``Reynolds--averaged'') Navier--Stokes and Boussinesq equations, including nonlinear advection and a quadratic feedback from the fluctuation field. In contrast, the evolution equation for the fluctuations is a modified version of the Swift--Hohenberg (SH) equation $\partial_t\psi = r\psi -(1+\nabla^2)^2\psi +\mathcal{N}(\psi)$ for the scalar field $\psi(x,z,t)$, where $\mathcal{N}$ is a nonlinear operator. The bifurcation parameter $r$ that multiplies the first term on the right-hand side has been replaced by the mean field $U(x,z,t)$, which here multiplies the second-order horizontal derivative of $\eta$. The inclusion of an SH-like operator ensures that the fast dynamics can experience exponential growth, depending upon the functional form and magnitude of the slow field $U$, thereby crudely mimicking linear instability of a stratified shear flow profile (for example). [Since $U$ multiplies the second-order horizontal derivative, the horizontal wavenumber of the fastest-growing instability also will depend on $U$---and, hence, may be expected to slowly evolve whenever $U$ does so.] The specific choice of the factors of $\varepsilon$ in (\ref{Nonlin_eta}) ensures that $\eta$ can, in principle, evolve on temporal and horizontal spatial scales that are a factor $\varepsilon$ smaller than the corresponding characteristic scales of evolution of the slow field $U$.

\subsection{Multiple scales analysis and QL reduction}

Owing to the scale separation induced by the small parameter $\varepsilon$, multiple scales analysis can be used to derive a reduced system. Let $\widetilde{T_s}$ and $\widetilde{X_s}$ denote the dimensional temporal and spatial scales over which slow processes occur (e.g., measured in days and kilometers) and $\widetilde{T_f}$ and $\widetilde{X_f}$ denote the corresponding characteristic dimensional scales for the fast processes (e.g., measured in seconds and meters). Then the dimensional time variable and horizontal spatial coordinate (also decorated with tildes) respectively satisfy
\begin{align}
    &\widetilde{t} = \tau \widetilde{T_f} = T\widetilde{T_s} \text{,}\\
    &\widetilde{x} = \chi \widetilde{X_f} = X\widetilde{X_s} \text{,}
\end{align}
where $\tau$ and $T$ are dimensionless `fast' and `slow' time variables and, similarly, $\chi$ and $X$ are dimensionless fast and slow horizontal coordinates. The scale separation between the fast and slow dynamics is measured by the small non-dimensional parameter $\varepsilon$: 
\begin{equation}
    \varepsilon = \dfrac{\widetilde{T_f}}{\widetilde{T_s}} = \dfrac{T}{\tau} \quad \text{and} \quad  \varepsilon = \dfrac{\widetilde{X_f}}{\widetilde{X_s}} = \dfrac{X}{\chi} \text{.}
\end{equation} 
Accordingly, to proceed with the analysis, we replace the single, original time variable $t$ with the \emph{two} time variables $\tau$ and $T$ and the single, original horizontal coordinate $x$ with the two coordinates $\chi$ and $X$, where 
\begin{center}
$t \rightarrow (T, \tau)$ \hspace{1cm} $T=t$ and $\tau =T/\varepsilon $,\\
$x \rightarrow (X, \chi)$ \hspace{1cm} $X=x$ and $\chi = X/\varepsilon\text{.} $
\end{center}
Crucially, in the analysis, $\tau$ and $T$ and, similarly, $\chi$ and $X$ are allowed to vary independently.

Allowing both $U$ and $\eta$ to depend on this expanded set of independent variables, we then posit the following asymptotic expansions: 
\begin{equation}
\label{multiscale_exp}
\begin{aligned}
&U=U_0+\varepsilon U_1 + \varepsilon^2 U_2 + O(\varepsilon^3),\\
&\eta=\eta_0+\varepsilon \eta_1 + \varepsilon^2 \eta_2 + O(\varepsilon^3),\\
&F=F_0+\varepsilon F_1 + \varepsilon^2 F_2 + O(\varepsilon^3).
\end{aligned}
\end{equation}
The system (\ref{Nonlin_U})--(\ref{Nonlin_eta}) can be solved order by order in $\varepsilon$ to obtain the leading order dynamics of the two fields.

Making use of the chain rule $\partial_t = \partial_T + \varepsilon^{-1}\partial_{\tau}$ and $\partial_x = \partial_X + \varepsilon^{-1}\partial_{\chi}$, and substituting (\ref{multiscale_exp}) into (\ref{Nonlin_U}), we immediately conclude that $U$ is independent of the fast spatial coordinate $\chi$ by considering (\ref{Nonlin_U}) at $O(\varepsilon^{-2})$, recalling the periodic boundary conditions in $\chi$:  
\begin{equation}
    \label{U_order_eps-2}
    \pdd{U_0}{\chi}=0.
\end{equation}
At $O(\varepsilon^{-1})$, the same equation confirms that $U_0$ is independent of the fast time variable $\tau$, since
\begin{equation}
    \label{U_order_eps-1}
    \pd{U_0}{\tau}=D\overline{\frac{\partial^2 U_1}{\partial\chi^2}}^{\chi}=0\text{,}
\end{equation}
where $\overline{(\cdot)}^{\chi}$ denotes a fast horizontal spatial average. 
Thus, $U_0 = U_0(X,z,T)$ and $U_1 = U_1(X,z,\tau, T)$.

The evolution equation for the slow variable is then obtained at $O(1)$:
\begin{multline}
\label{multiscale_U_O1}
\pd{U_0}{T}  +\pd{U_1}{\tau}+ U_0\pd{U_0}{X} +  U_0\pd{U_1}{\chi} + U_1\pd{U_0}{\chi} \\= F -\nu U_0 + D\pdd{U_0}{z} - \left(\pd{\eta_0}{\chi}\right)^2 \text{ .}
\end{multline}

\noindent
Following \citet{Michel2019} and \citet{Chini2022}, we introduce a fast averaging procedure over both the fast time $\tau$ and spatial coordinate $\chi$, defined for a generic function $\psi$ as
\begin{multline}
\label{fastav_tauchi}
\overline{\psi}(X, z, t; \varepsilon)\\=\lim_{\tau_{f}, L{\chi_f}\to\infty} \frac{1}{\tau_f}\frac{1}{L_{\chi_f}}\int_0^{\tau_f} \int_0^{L_{\chi_f}}\psi (\chi, X, z, \tau, t; \varepsilon)d\tau d\chi\text{,}
\end{multline}
where $L{\chi_f}$ and $\tau_{f}$ are dimensionless spatial and temporal scales intermediate in scale to the domains over which $X$, $\chi$ and $T$, $\tau$ are defined, respectively.
Using (\ref{fastav_tauchi}), the fast-average of (\ref{multiscale_U_O1}) yields the evolution equation for the mean variable $U_0(X, z, T)$: 
\begin{equation}
\label{U_GQL}
\pd{U_0}{T}  + U_0\pd{U_0}{X}  = F -\nu U_0 + D\pdd{U_0}{z} - \overline{\left(\pd{\eta_0}{ \chi}\right)^2}\text{ .}
\end{equation}

Similarly, the dynamics of the leading-order fast-varying field $\eta_0$ is obtained from (\ref{Nonlin_eta}), after dividing through by $\varepsilon$, at $O(\varepsilon^{-1})$:
\begin{equation}
\label{eta_GQL}
\pd{\eta_0}{\tau} = -\eta_0 - U_0\pdd{\eta_0}{\chi} - \dfrac{\partial^4 \eta_0}{\partial \chi^4} + \pdd{\eta_0}{z} \text{.}
\end{equation}
 
By inspection of (\ref{eta_GQL}), the fast dynamics is homogeneous and linear in the fluctuations, and directly coupled to the mean dynamics. In particular, the evolution of the fluctuation field $\eta_0$ is parameterically dependent upon the mean field, as $U_0$ is a prefactor in the second-derivative term on the right-hand-side of (\ref{eta_GQL}) that drives instability. In turn, the fluctuation dynamics modify the evolution of the mean field via the `Reynolds-stress divergence' feedback term $\overline{\partial_\chi^2\eta_0}$ in (\ref{U_GQL}).

Collectively, the above attributes imply that the system (\ref{U_GQL})--(\ref{eta_GQL}) falls in the category of \textit{generalised} QL (GQL) models \cite{Marston2016}, which differ from strict QL models via the presence of the mean/mean nonlinearity in the mean field equation. This latter attribute, represented here by the advective term in (\ref{U_GQL}), allows for the interaction between low Fourier modes, turning the strict horizontal mean in a QL reduction of the dynamics into a slowly spatially-varying mean. Although of fundamental importance in many physical systems, including strongly stratified shear flow turbulence, the dependence of $U$ and $\eta$ on the slow coordinate $X$ (and thus the presence of the advective term in (\ref{U_GQL})) will be suppressed henceforth 
to simplify the analysis. Upon re-interpreting the spatial part of the fast averaging operation in (\ref{fastav_tauchi}) as an average over the entire domain in $\chi$, a strict QL formulation of (\ref{U_GQL}) reads
\begin{equation}
\pd{U_0}{T} = F -\nu U_0 + D\pdd{U_0}{z} - \overline{\left(\pd{\eta_0}{ \chi}\right)^2}\text{.}
    \label{U_QL}
\end{equation}
The linearity and the autonomy in $\chi$ and $\tau$ of the fluctuation equation allows for the general solution of (\ref{eta_GQL}) to be expressed as a superposition of modal solutions of the form
\begin{equation}
\label{ansatz}
\eta_0 = A\hetaO(z)e^{\sigma(k)\tau}e^{ik\chi} + c.c.\, ,
\end{equation}
where $c.c.$ denotes `complex conjugate', and the real amplitude factor $A$, the wavenumber $k$, the wavenumber-dependent, possibly complex growth rate $\sigma(k)$, and the vertical eigenfunction $\hat{\eta}(z)$ all can vary slowly in $T$ owing to the potential variation of the mean field $U_0$ with $T$.

At this stage of the derivation, it is crucial to notice that the leading-order fast dynamics lacks an \emph{explicit} saturation mechanism. Instead, saturation of growing fluctuations can be achieved only through the flux feedback in (\ref{U_GQL}). In the limit $\varepsilon\to 0$, this feedback is only possible in a state of marginal stability. That is, negative growth rates would cause an exponentially fast damping of the fluctuations on the fast time scale, corresponding to their instantaneous decay on the slow time scale. Conversely, positive growth rates are not compatible with a convergent averaged fluctuation feedback, thereby breaking the posited scale separation. While the first case, $\Re\{\sigma\}<0$, does not invalidate the QL reduction, allowing for a filtered dynamics in which the slow field evolves in the absence of any collective effect of the fluctuations, the second case, $\Re\{\sigma\}>0$, requires the restoration of the nonlinear terms in the fast dynamics and the co-evolution of the fast and slow fields until a marginally-stable state is reestablished.

Omitting the subscript $0$ for brevity of notation and substituting the ansatz (\ref{ansatz}) into (\ref{U_QL}) and (\ref{eta_GQL}), the simplified, reduced dynamics is described by an initial-value problem for the mean field $U$ and an eigenvalue problem for the fast modes $\heta$:
\begin{equation}
\label{U_QL_IVP}
\pd{U}{T} = F(z,T) -\nu U -2K^2|A|^2|\hphi|^2+ D \pdd{U}{z}\text{,}
\end{equation}

\begin{equation}
\label{eta_QL_EVP}
\sigma \heta =\left(-1 +k^2U -k^4 +\pdd{}{z}\right)\heta \equiv \Lop\heta \text{ .}
\end{equation}
\noindent
For clarity of the subsequent exposition, we have denoted by $\heta$ the eigenfunctions associated with the eigenvalue problem (\ref{eta_QL_EVP}), in principle defined for \emph{any} growth rate, and by $\hphi$ the marginally stable (leading) fast mode that feeds back on the slow field only in a condition of marginal stability. Thus, $\hphi \equiv \heta$ when $\Re\{\sigma \} = 0$.
Analogously, we have used $K$ to denote the wavenumber associated with the neutral mode $\hphi$ in (\ref{U_QL_IVP}) to stress the independence of the mean field evolution on $k$. Finally, note that the  marginal stability requirement (i.e., $\Re\{\sigma\}=0$),  necessary for the feedback term in (\ref{U_QL_IVP}) to be finite, has been accounted for by substituting a zero growth rate while performing the fast-averaging over $\tau$ and $\chi$.

\subsection{Marginal stability constraint: Slaving of the fluctuation amplitude}
\label{sec:slavingA}

The coupling of the fast and slow components of the system (\ref{Nonlin_U})--(\ref{Nonlin_eta}) in a condition of marginal stability occurs via the modal amplitude $A(T)$, which is \emph{a priori} unknown. \citet{Michel2019} showed that for slow--fast QL systems subject to fast instabilities, an amplitude equation \emph{cannot} be derived via the usual protocol of insuring the solvability of the equations governing the higher-order fluctuation fields, owing to the lack of closure of the perturbative equation hierarchy. These authors demonstrated that a suitable constraint to determine the amplitude $A(T)$ can be derived by instead exploiting the self-tuning of the slow dynamics toward a marginally stable state. In particular, when the leading eigenvalue approaches zero (from below), the amplitude must be algebraically slaved to the mean field in order to maintain its marginal stability. 

\begin{figure}[ht]
\centering
\includegraphics[width=0.45\textwidth]{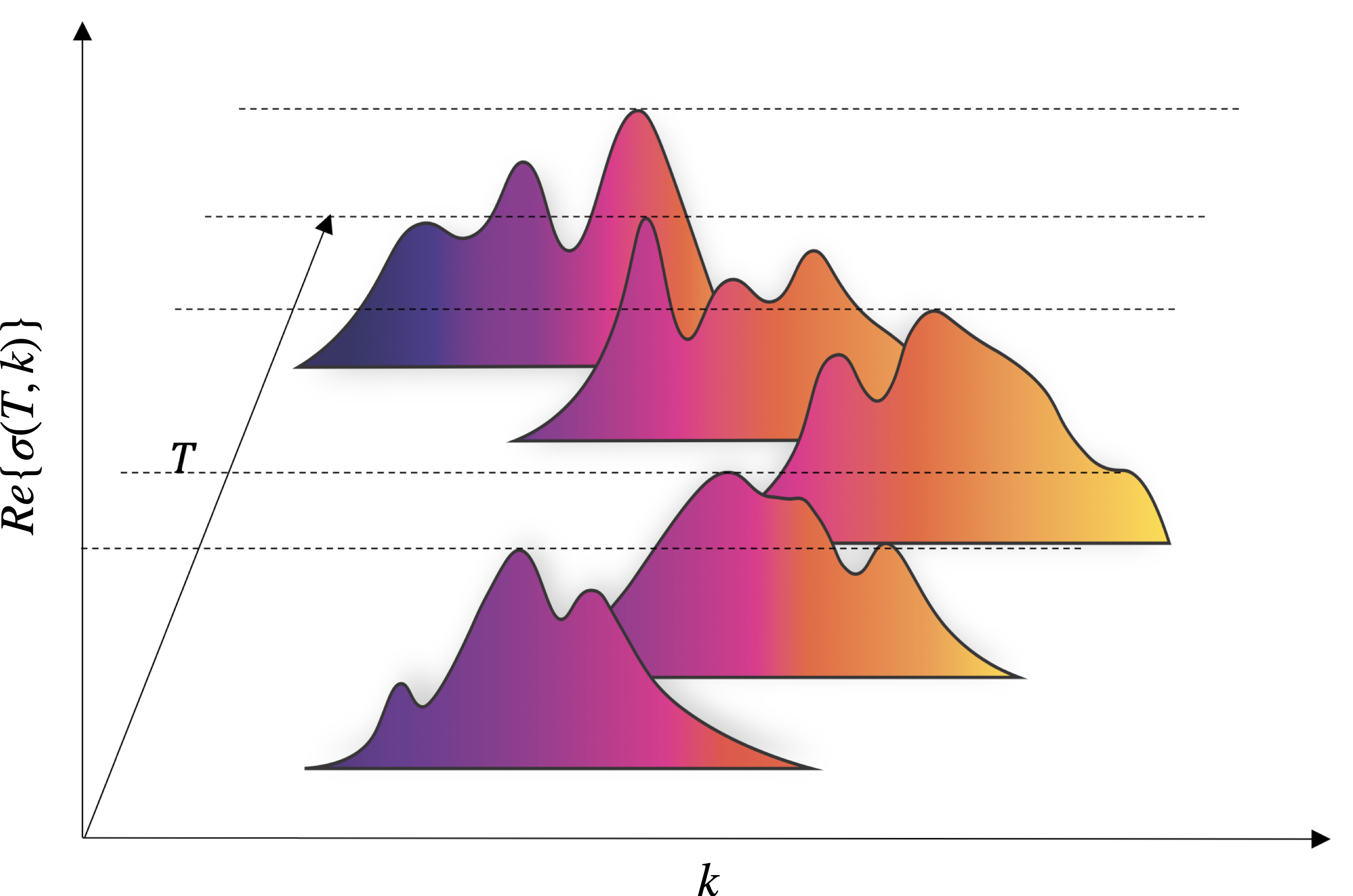}
\caption{Schematic of the time-dependent dispersion relation $\Re{\sigma(T,k)}$. Black dashed lines represent the growth rate of the of the fastest growing mode at a fixed time $T$ for which $\Re\{\partial_k\sigma\}=0$.}
\label{fig:sigma_Tk}
\end{figure}

This constraint is obtained by first deducing a slow evolution equation not for the amplitude but for the leading eigenvalue (growth rate) $\sigma$, \emph{viz.}, $\Re\{\partial_T\sigma\}$, which can be obtained as a first order correction of the time-perturbed eigenvalue problem (\ref{eta_QL_EVP}). In 2D, however, the introduction of the `additional' spatial coordinate $\chi$ implies that the eigenvalue problem depends not only on $T$ through the mean field but also, at fixed $T$, on the wavenumber $k$. Consequently, a second constraint is required to ensure the marginal stability condition holds over all admissible wavenumbers. As suggested in the schematic in figure~\ref{fig:sigma_Tk}, one approach \cite{Chini2022} is simply to identify the wavenumber $K$ of the fastest-growing mode by a search over $k$ at fixed $T$ and, subsequently, to slave the amplitude of the corresponding mode (with fixed wavenumber $K$) to the mean field.

The key idea of this approach is to exploit the natural tendency of the slow dynamics to evolve near to a state of marginal stability, as noted above, in order to prescribe the fluctuation amplitude required to maintain that state. Consequently, the slow time derivative of the growth rate  has to be zero whenever the growth rate $\Re\{\sigma\} = 0$. To derive the necessary condition, we employ a first-order perturbation analysis. Expanding the eigenvalue problem (\ref{eta_QL_EVP}) in time at fixed $k=K$ around a point $M$ at which the growth rate is maximum (and $T=T_M$) yields
\begin{equation}
    \begin{aligned}
    \label{pert_sigma_1st}
    \sigma(T_M +\Delta T,K) & \sim\sigma_M + \pd{\sigma}{T}\EvalM\Delta T, 
\end{aligned}
\end{equation}
\begin{equation}
    \begin{aligned}
    \label{pert_L_1st}
    \Lop(T_M +\Delta T,K) & \sim\Lop_M + \pd{\Lop}{T}\EvalM\Delta T,
\end{aligned}
\end{equation}
\begin{equation}
\begin{aligned}
    \label{pert_eta_1st}
    \heta(T_M +\Delta T,K) & \sim\heta_M + \pd{\heta}{T}\EvalM\Delta T.
\end{aligned}
\end{equation}
We then collect terms order by order in the arbitrarily small time increment $\Delta T$, and obtain the following boundary-value problem at $O(\Delta T)$:
\begin{equation}
    \label{BVP_DT}
    \Lopc_{M} \pd{\heta}{T}\EvalM = -\pd{\Lop}{T}\EvalM \heta_M + \pd{\sigma}{T}\EvalM \heta_M,
\end{equation}  

where the operator $\Lopc$ is defined as $\Lopc = \Lop -\sigma$ and $\Lopc_M$ indicates evaluation at point $M$.
Because $\Lopc$ is a singular operator (by construction), the existence of solutions of (\ref{BVP_DT}) has to be ensured by imposing a solvability condition.

Defining the $L^2$ inner product 
\begin{equation}
    \bra{\psi_1}\ket{\psi_2}=\int_0^L \psi_1(z)\psi_2^*(z)dz 
\end{equation}
for two functions $\psi_1(z)$ and $\psi_2(z)$,
the Fredholm Alternative theorem states, in the context of singular boundary value problems, that the generic problem $\Lopc u = f $ (equipped with suitable boundary conditions) is solvable if and only if the right-hand-side $f$ is orthogonal to the null space of the adjoint linear operator $\Lopc ^{\dagger}$. Thus, denoting by $\psi_n$ and $\psi_n^{\dagger}$ (for integer $n$) the eigenvectors of the direct and adjoint operators respectively associated with the eigenvalues $\lambda_n$ and $\lambda_n^*$, the solvability of $\Lopc u = f$ requires
\begin{equation}
    \label{solvab}
    \bra{f}\ket{\psi_0^{\dagger}}=\bra{\Lopc u}\ket{\psi_0^{\dagger}}=\bra{u}\ket{\Lopc^{\dagger}\psi_0^{\dagger}}=0;
\end{equation}
the direct problem then has infinitely many solutions, 
\begin{equation}
    u=C\psi_0 + \sum_{n=1}^{\infty} \dfrac{\bra{f}\ket{\psi_n^{\dagger}}}{\lambda_n\bra{\psi_n}\ket{\psi_n^{\dagger}}}\psi_n,
\end{equation}
for arbitrary constant $C$. 

By inspection of (\ref{eta_QL_EVP}), we observe that the linear operator $\Lop$ is self-adjoint, implying that its eigenvalues are real and the corresponding eigenvectors can be chosen real. Imposing the solvability condition (\ref{solvab}) on the boundary value problem (\ref{BVP_DT}) then yields
\begin{equation}
    \label{solvab_DT}
    \bra{\Lopc_M \pd{\heta}{T}\EvalM} \ket{\heta_M^{\dagger}}=\bra{\pd{\heta}{T}\EvalM}\ket{\Lopc_M^{\dagger}\heta_M^{\dagger}}=0.
\end{equation}
Exploiting the self-adjoint nature of the operator $\Lopc$ in this specific problem, we obtain\\
\begin{multline}
\label{solveable1}
    \bra{\Lopc_M \pd{\heta}{T}\EvalM} \ket{\heta_M}=-\intz{\bigg(\pd{L}{T}\EvalM\bigg)|\heta_M|^2} \\+ \pd{\sigma}{T}\EvalM\intz{|\heta_M|^2} = 0,
\end{multline}
where 
\begin{multline}
\label{dLdT}
    \pd{\Lop}{T} = k^2\pd{U}{T} = k^2 \bigg(F-\nu U + D\pdd{U}{z}\bigg) -2k^2|A|^2K^2|\hphi|^2
\end{multline}
 and 
\begin{equation}
\label{dLdT_M}
    \pd{\Lop}{T}\EvalM = K^2\bigg(F-\nu U + D\pdd{U}{z}\bigg)-2|A|^2K^4|\hphi|^2, 
\end{equation}
given that $k = K$ when $T=T_M$.
Adopting the following arbitrary normalization condition for the eigenfunctions,  
\begin{equation}
\label{normalization}
    \braket{\heta}=\int_{0}^{L_z}|\heta|^2dz=1,
\end{equation}
the solvability condition (\ref{solveable1}) directly yields an evolution equation for the growth rate $\sigma$ in slow time:
\begin{equation}
    \pd{\sigma}{T}\EvalM=K^2\intz{\bigg(\pd{U}{T}\bigg)|\heta_M|^2} \text{ .}
\end{equation}
Finally, substituting the evolution equation (\ref{U_QL_IVP}) for the mean field yields
\begin{multline}
\label{dt_sigma}
    \pd{\sigma}{T}\EvalM=K^2\intz{\bigg(F-\nu U + D\pdd{U}{z}\bigg)|\heta_M|^2}\\
    -2K^4|A|^2\intz{|\heta_M|^2|\hphi|^2} \text{ .}
\end{multline}
An expression for the fluctuation amplitude then can be derived by imposing the marginal stability constraint $\partial_T \Re{\sigma}|_M = \partial_T \sigma|_M = 0$ when $\sigma = 0$, \emph{viz.},
\begin{equation}
\label{2D_Amplitude}
    |A(T)|=\sqrt{\frac{\alpha}{2K^2\beta}} \text{ ,}
\end{equation}
where
\begin{equation}
 \label{2D_alpha}
 \alpha = \intz{\left(F-\nu U - D\pdd{U}{z}\right)|\hphi|^2} \text{ ,}
\end{equation}
\begin{equation}
 \label{2D_beta}
 \beta=\intz{|\hphi|^4}\text{ .}
\end{equation}

Compactly re-writing (\ref{dt_sigma}) as
\begin{equation}
\pd{\sigma}{T}\EvalM=K^2\alpha -2K^4|A|^2\beta ,   
\end{equation}
it can be noted that the sign of the integrals $\alpha$ and $\beta$ determine whether or not an amplitude $|A|$ exists (when marginal stability is attained). Positive values of $\alpha$ correspond to linear processes driving instability growth while positive values of $\beta$ ensure the nonlinear feedback from the fluctuations is stabilizing. In contrast, negative values of $\alpha$ drive the system toward more stable states ($\partial_T\sigma<0$), while negative values of $\beta$ in principle would (if $\alpha>0$) induce potentially large-amplitude `bursting' events ($\partial_T\sigma>0$) that generally would break the posited scale separation. Although the fluctuation feedback generally is not sign-definite, in the model problem under consideration $\beta$ is positive-definite, precluding the occurrence of such bursting events.

\section{prediction of the fastest growing mode}
\label{sec_kpred}
As discussed in \S~\ref{sec:intro}, imposition of the marginal stability constraint for 2D systems possessing a spatially extended coordinate direction (parameterized by a wavenumber $k$) is considerably more involved than in one-dimensional (1D) systems, e.g., as considered by \citet{Michel2019}, due to the need to ensure that at each slow time instant the marginal mode is \emph{simultaneously} the fastest-growing mode over all $k$. \citet{Chini2022} addressed this additional requirement by solving a sequence of eigenvalue problems for a range of $k$ values to locate the fastest-growing mode and then employing the algorithm described in \S~\ref{sec:slavingA} to determine the amplitude of that mode. Not only is this approach inelegant, but it is also computationally costly, given the need to repeatedly solve 1D eigenvalue problems at each slow time instant. In this section, we derive an evolution equation for the wavenumber of the fastest-growing, marginal mode that obviates the need for a brute-force search over an effectively continuous range of wavenumbers $k$. This equation is obtained by simultaneously enforcing the tangency conditions

\begin{equation}
    \partial_T\Re\{\sigma\}\eval_{(k_M,T_M)}=0 \quad \land \quad \partial_k\Re\{\sigma\}\eval_{(k_M,T_M)}=0,
\end{equation}
where we have introduced the notation $k_M = K(T_M)$. That is, $k_M$ denotes the wavenumber of a mode satisfying \emph{both} tangency conditions while
$K(T)$ remains a differentiable function of time $T$ and indicates the wavenumber of the mode with maximum growth rate at any given time.

\subsection{Evolution equation for the wavenumber of the fastest-growing, marginal mode}
\begin{figure}[htbp]
\centering
\includegraphics[width=0.5\textwidth]{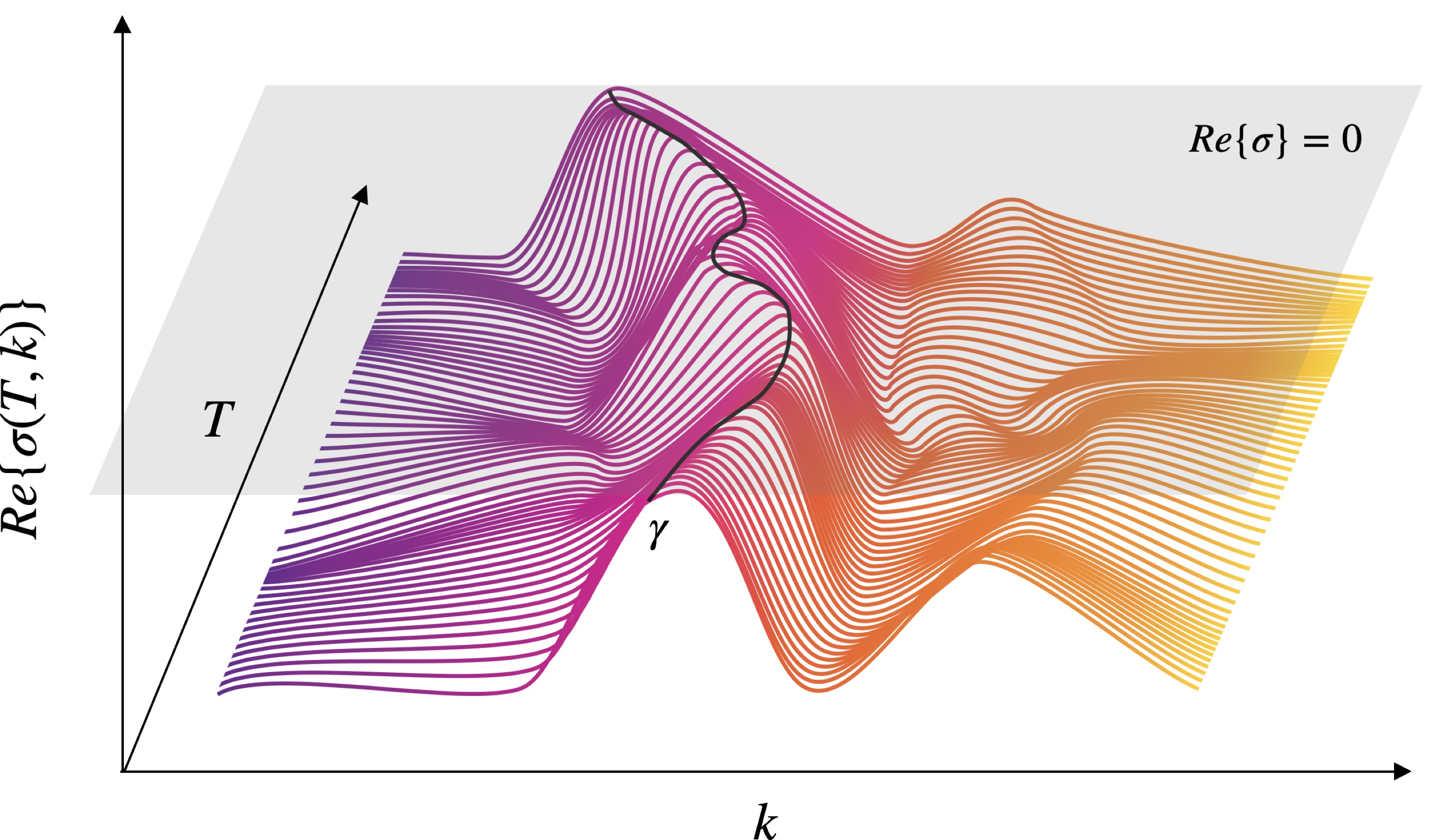}
\caption{Schematic of the time-dependent dispersion relation $\Re{\sigma(T,k)}$ as a continuous surface in $\reals^3$. The black line represents the curve $\gamma$, obtained from the intersection between the surface $\Gamma$ and the horizontal $k$--$T$ plane at $\Re{\sigma}=0$.  }
\label{fig:sigma_landscape}
\end{figure}

From a geometrical point of view, the time-dependent dispersion relation (for a generic operator) $\Re{\sigma(T,k)}$, illustrated in figure \ref{fig:sigma_landscape}, is represented by a three-dimensional landscape that can be described as a continuous parameterized surface in $\reals^3$.
Defining a continuous mapping function $\Gamma: A\rightarrow \reals^3$, where $A\subseteq \reals^2$, a generic parameterization of a surface is given by 
\begin{equation}
\Gamma(T,k)=(\Gamma_1(T,k), \Gamma_2(T,k), \Gamma_3(T,k)).
\end{equation}
By construction, however, $\Gamma$ is a locally injective map, enabling the specific choice for the parametric equations $\Gamma_1(T,k)=T$, $\Gamma_2(T,k)=k$ to be made. Thus, the surface can be described as the graph
\begin{equation}
    \label{Gamma_graph}
    \Gamma(T,k)=(T,k,\Re{\sigma(T,k)}).
\end{equation}

Despite the operator $L$ being self-adjoint for the given model problem (\ref{eta_QL_EVP})(hence $Re\{\sigma\}=\sigma$), to keep the formulation generic we reintroduce the detailed notation for the real and imaginary part of $\sigma$ and its derivatives, discarding it solely in the context of the model problem evaluation.
For brevity of notation, the real and the imaginary part will be denoted using the subscripts $r$ and $i$, respectively:
\begin{equation}
    \partial_T^n\partial_k^m\sigma_r =\partial_T^n\partial_k^m\Re{\sigma} \quad \text{and}\quad \partial_T^n\partial_k^n\sigma_i =\partial_T^n\partial_k^m\Im{\sigma},
\end{equation}
for $n,m = 0, 1, 2, 3,\ldots$
The marginal stability requirements to be satisfied by the fastest growing mode at each time results in the presence of a ridge (a curve in $\reals^3$) on the surface $\Gamma$, that lies in the horizontal plane $(T,k,\sigma_r=0)$ and is characterized by $\nabla\sigma_r=\mathbf{0}$ at each point, where here $\nabla = (\partial_T,\partial_k)$.

Formally, a parametrised curve $\gamma$ on a surface $\Gamma$ is described as a pair of smooth curves $\mu$ and $\gamma$ satisfying $\gamma=\Gamma \circ \mu$. The plane curve $\mu$ is said to be the coordinate curve of the pair and here it is the projection of $\gamma$ onto the two-dimensional space spanned by $T,k$:
\begin{equation}
    \mu(T)=(T,K(T)),
\end{equation}
where $\mu: I\rightarrow\reals^2$, with $I\subseteq \reals$, and $K(T)$ is a smooth function $K:I\rightarrow\reals$.
The function $\gamma$, describing the ridge, is then uniquely determined by the composition $\gamma=\Gamma \circ \mu$; \emph{viz.},
\begin{equation}
\label{2D_ridge}
    \gamma(T)=(T,K(T),\sigma_r(T,K(T))) = (T, K(T), 0),
\end{equation}
given that the ridge is the intersection between the three-dimensional landscape and the plane $\sigma_r=0$.

It follows that, at each time $T$, the parabolic point $M$ on the ridge $\gamma$ that satisfies 
\begin{equation}
\label{parab_conditions}
\begin{aligned}
    &\sigma_r(T_M,k_M)=0 \text{,}\\
    &\partial_k\sigma_r|_{(T_M,k_M)}=0 \text{,}\\
    &\partial_T\sigma_r|_{(T_M,k_M)}=0 \text{ .}
\end{aligned}
\end{equation}

Within this framework, the prediction of $k$ in a state of marginal stability requires the derivation of an evolution equation for the function $K(T)$ enforcing the propagation in time of the parabolic point properties (\ref{parab_conditions}) along $\gamma$. 

Letting $M=(T_M, k_M)$ denote a point in the coordinate space indicating the location of the ridge and $\bar{\delta}$ the tangent vector to the coordinate curve $\mu$ ,
\begin{equation}
\bar{\delta}=\mu'=\bigg(1,\td{K(T)}{T}\bigg),
\end{equation}
the Taylor series expansion of the growth rate around $M$ \emph{along} the ridge reads
\begin{widetext}
\begin{equation}
\label{2D_Taylor_ridge}
\begin{aligned}
     \sigma_r ((T,k)_M +\Delta \Bar{\delta}) &\sim\sigma_r (T_M,k_M) 
     + \bigg(\pd{\sigma_r}{T}\EvalM +\pd{\sigma_r}{k}\EvalM \td{K(T)}{T}\bigg)\Delta \\
      & + \frac{1}{2}\bigg( \pdd{\sigma_r}{T}\EvalM + \pdd{\sigma_r}{k}\EvalM\bigg(\td{K(T)}{T}\bigg)^2 + 2\pdm{\sigma_r}{T}{k}\EvalM\td{K(T)}{T}\bigg)\Delta^2 +O(\Delta^3)\\
      & = \frac{1}{2}\bigg( \pdd{\sigma_r}{T}\EvalM + \pdd{\sigma_r}{k}\EvalM\bigg(\td{K(T)}{T}\bigg)^2 + 2\pdm{\sigma_r}{T}{k}\EvalM\td{K(T)}{T}\bigg)\Delta^2 +O(\Delta^3),
\end{aligned}
\end{equation}
\end{widetext}
where $\Delta\ll 1$ and the marginal stability condition at $M$ and the parabolic point conditions (\ref{parab_conditions}) have been imposed in the last line of (\ref{2D_Taylor_ridge}).
Requiring then the preservation of marginal stability along $\mu$ in (\ref{2D_Taylor_ridge}), i.e. $\sigma_r((T,k)_M+\Delta \Bar{\delta})=0$, or equivalently a zero Gaussian curvature of the surface $\Gamma$ at the point $M$, yields an evolution equation for $K(T)$:
\begin{equation}
\label{2D_dKdT2}
    \bigg(\td{K}{T}\bigg)^2  + 2\frac{\partial_k(\partial_T\sigma_r)\evalM}{\partial_k^2\sigma_r\evalM}\bigg(\td{K}{T}\bigg)= -\frac{\partial_T^2\sigma_r\evalM}{\partial_k^2\sigma_r\evalM}.
\end{equation}

We stress that, while the eigenvalue problem for the fast modes (\ref{eta_QL_EVP}) is well defined at any point on the surface $\Gamma$ and therefore for any wavenumber $k$, the fluctuation feedback on the slow dynamics (\ref{U_QL_IVP}) is given only by the marginally stable eigenmodes on the ridge $\gamma$, for which the wavenumber $k=K(T)$ is defined as a function of time, yielding 
\begin{equation}
   \pd{L}{k}\EvalM=\bigg(2kU-4k^3\bigg)\EvalM=2KU-4K^3 \text{ .}
\end{equation}

It follows that the neutral eigenfunctions $\hphi$ are only functions of space and time and do not depend on the wavenumber $k$: $\hphi=\hphi(z,T)$. Thus, while the evaluation of the generic eigenfunction $\heta$ at a point $M$ on the ridge is exactly $\hphi$,
\begin{equation}
\label{2D_etaphi}
    \heta_M = \heta(z,T,k=K(T)) = \hphi(z,T) \text{ ,}
\end{equation}
the partial derivatives of $\hphi$ and $\heta$ (evaluated at $M$) with respect to $T$ are different:


\begin{equation}
\label{2D_dT_etaphi}
\pd{\hphi}{T} = \pd{\heta_M}{T}=\pd{\heta}{T}\EvalM + \pd{\heta}{k}\EvalM\td{K}{T} \text{.}
\end{equation}

As for the determination of the fluctuation amplitude, the prediction of the corresponding wavenumber via (\ref{2D_dKdT2}) requires specification of the first and here also the second partial derivatives of the growth rate $\sigma_r(k,T)$.
To construct these derivatives, we again employ perturbation analysis. Infinitesimally perturbing the eigenvalue problem (\ref{eta_QL_EVP}) in $k$ and $T$ yields
\begin{widetext}   
\begin{equation}
    \begin{aligned}
    \label{pert_sigma}
    \sigma(T_M & +\Delta T,k_M+\Delta k) \sim\sigma_M + \pd{\sigma}{T}\EvalM\Delta T + \pd{\sigma}{k}\EvalM \Delta k + \frac{1}{2}\pdd{\sigma}{T}\EvalM\Delta T^2 + \frac{1}{2}\pdd{\sigma}{k}\EvalM \Delta k^2 +\pdm{\sigma}{k}{T}\EvalM\Delta T \Delta k,
\end{aligned}
\end{equation}

\begin{equation}
    \begin{aligned}
    \label{pert_L}
    \Lop(T_M & +\Delta T,k_M+\Delta k)  \sim\Lop_M + \pd{\Lop}{T}\EvalM\Delta T + \pd{\Lop}{k}\EvalM \Delta k + \frac{1}{2}\pdd{\Lop}{T}\EvalM\Delta T^2 + \frac{1}{2}\pdd{\Lop}{k}\EvalM \Delta k^2 +\pdm{\Lop}{k}{T}\EvalM\Delta T \Delta k,
\end{aligned}
\end{equation}

\begin{equation}
\begin{aligned}
    \label{pert_eta0}
    \heta(T_M & +\Delta T,k_M+\Delta k)  \sim\heta_M + \pd{\heta}{T}\EvalM\Delta T + \pd{\heta}{k}\EvalM \Delta k + \frac{1}{2}\pdd{\heta}{T}\EvalM\Delta T^2 + \frac{1}{2}\pdd{\heta}{k}\EvalM \Delta k^2 +\pdm{\heta}{k}{T}\EvalM\Delta T \Delta k.
\end{aligned}
\end{equation}

\vspace{0.3cm}

Collecting terms order by order, the following boundary values problems are obtained:\\  

 $\boldsymbol{O(\Delta T)}$
\begin{equation}
    \label{2D_BVP_DT}
    \Lopc_M \pd{\heta}{T}\EvalM = -\pd{\Lop}{T}\EvalM \heta_M + \pd{\sigma}{T}\EvalM \heta_M
\end{equation}  

$\boldsymbol{O(\Delta k)}$
\begin{equation}
    \label{2D_BVP_Dk}
    \Lopc_M \pd{\heta}{k}\EvalM = -\pd{\Lop}{k}\EvalM \heta_M + \pd{\sigma}{k}\EvalM \heta_M
\end{equation}\\

 $\boldsymbol{O(\Delta T \Delta k)}$
 \begin{equation}
    \label{2D_BVP_DTDk}
    \begin{aligned}
      \Lopc_M \pdm{\heta}{k}{T}\EvalM = & -\pdm{\Lop}{k}{T}\EvalM \heta_M  -\pd{\Lop}{k}\EvalM\pd{\heta}{T}\EvalM 
      -\pd{\Lop}{T}\EvalM\pd{\heta}{k}\EvalM + \pd{\sigma}{T}\EvalM\pd{\heta}{k}\EvalM 
      + \pd{\sigma}{k}\EvalM\pd{\heta}{T}\EvalM+  \pdm{\sigma}{k}{T}\EvalM \heta_M   
    \end{aligned}
\end{equation}\\

$\boldsymbol{O(\Delta T^2)}$
 \begin{equation}
    \label{2D_BVP_DT2}
    \begin{aligned}
      \Lopc_M \pdd{\heta}{T}\EvalM = & -\pdd{\Lop}{T}\EvalM \heta_M  -2\pd{\Lop}{T}\EvalM\pd{\heta}{T}\EvalM 
      + 2\pd{\sigma}{T}\EvalM\pd{\heta}{T}\EvalM+  \pdd{\sigma}{T}\EvalM \heta_M   
    \end{aligned}
\end{equation}\\

$\boldsymbol{O(\Delta k^2)}$
 \begin{equation}
    \label{2D_BVP_Dk2}
    \begin{aligned}
      \Lopc_M \pdd{\heta}{k}\EvalM = & -\pdd{\Lop}{k}\EvalM \heta_M  -2\pd{\Lop}{k}\EvalM\pd{\heta}{k}\EvalM 
       + 2\pd{\sigma}{k}\EvalM\pd{\heta}{k}\EvalM+  \pdd{\sigma}{k}\EvalM \heta_M   
    \end{aligned}
\end{equation}\\

\end{widetext}

Note that evaluation of the second partial derivatives of the growth rate $\partial_T^2\sigma$ and $\partial_k\partial_T\sigma$ requires the solution for the first-order eigenfunction correction $\partial_T\heta|_M$. As for the solution of a generic singular boundary value problem $\Lopc u = f$, this solution can be obtained using the generalized inverse computed, e.g., via a QR decomposition. Practically, one must solve
\begin{equation}
    \label{2D_genr_inverse}
    \Lopc u_p = f - \bra{f}\ket{\psi_0^{\dagger}}\psi_0^{\dagger},
\end{equation}
where $\psi^{\dagger}_0 \in ker(\Lopc^{\dagger})$ and $u_p$ is a particular solution of $\Lopc u = f$ that minimizes the least squares error 
\begin{equation}
    \label{2D_least_square_error}
    \norm{\Lopc u - f}^2 \text{ .}
\end{equation}
The general solution of the system is then defined up to an arbitrary multiple of the null eigenfunction $\psi_0$ (where $\Lopc\psi_0=0$): 
\begin{equation}
    u = C\psi_0 + u_p,
\end{equation}
from which a specific solution can be selected by imposing the orthogonality of $u$ with respect to $\psi_0$, $\bra{u}\ket{\psi_0} = 0$, which gives
\begin{equation}
    C = -\dfrac{\bra{u_p}\ket{\psi_0}}{\braket{\psi_0}} \text{ .}
\end{equation}

Returning to the boundary-value problem (\ref{2D_BVP_DT}) at order $\Delta T$, we obtain for the first order correction of the eigenfunction 
\begin{equation}
    \label{2D_etaT}
    \pd{\heta}{T}\EvalM = C_1\heta_M + \bigg(\pd{\heta}{T}\bigg)_p\EvalM,
\end{equation}
with the multiplicative constant
\begin{equation}
    C_1 = -\bra{\bigg(\pd{\heta}{T}\bigg)_p\EvalM}\ket{\heta_M},
\end{equation}
where the normalization condition (\ref{normalization}) has been imposed. The orthogonality between the homogeneous solution $\eta_M$ and the solution $\partial_T\heta|_M$ is, in this case, equivalent to enforcing the preservation of the norm 
\begin{equation}
    \label{2D_norm_preservation}
    \pd{}{T}\bra{\heta_M}\ket{\heta_M} = 2\intz{\pd{\heta}{T}\EvalM\heta_M} =0\text{ .}
\end{equation}

By applying this procedure, namely, 
\begin{enumerate}
    \item determining the partial derivative of the growth rate via Fredholm alternative, 
    \item imposing the parabolic-point properties in a state of marginal stability, and
    \item computing the correction of the eigenfunction via the generalized inverse algorithm and preservation of the norm,
\end{enumerate}
to the boundary-value problem (\ref{2D_BVP_DT}) at order $O(\Delta k)$, we obtain the following results.\\

\noindent
$\boldsymbol{O(\Delta k)}$
\begin{equation}
    \pd{\sigma}{k}\EvalM = 2K\intz{U|\heta_M|^2}-4K^3,
\end{equation}
which, after imposing the maximum growth-rate constraint $\partial_k\sigma|_M = 0$, yields the integral relation 
\begin{equation}
2K^2 = \intz{U|\heta_M|^2}
\end{equation}
and thence
\begin{equation}
    \label{2D_etak}
    \pd{\heta}{k}\EvalM = C_2\heta_M + \bigg(\pd{\heta}{k}\bigg)_p\EvalM,
\end{equation}
 with 
\begin{equation}
    C_2 = -\bra{\bigg(\pd{\heta}{k}\bigg)_p\EvalM}\ket{\heta_M}.
\end{equation}

\begin{widetext}
Finally, applying steps~1. and~2.~to the boundary value problems (\ref{2D_BVP_DTDk})--(\ref{2D_BVP_Dk2}) at second order gives\\ 

$\boldsymbol{O(\Delta k)^2}$
\begin{equation}
    \label{2D_sigmakk}
    \pdd{\sigma}{k}\EvalM = -8K^2 + 4K\intz{U\heta_M\pd{\heta}{k}\EvalM}
\end{equation}

$\boldsymbol{O(\Delta T)^2}$
\begin{equation}
    \label{2D_sigmaTT}
    \begin{aligned}
        \pdd{\sigma}{T}\EvalM = -2K^2\td{K}{T}\bigg(\intz{(F-\nu U + D\partial_z^2U)\heta_M\pd{\heta}{k}\EvalM} -\dfrac{\alpha}{\beta}\intz{\heta_M^3\pd{\heta}{k}\EvalM} \bigg)
    \end{aligned}
\end{equation}

$\boldsymbol{O(\Delta T \Delta k)}$
\begin{equation}
\label{2D_sigmakT}
\begin{aligned}
    \pdm{\sigma}{k}{T}\EvalM = &\;2K\intz{U\heta_M\pd{\heta}{T}\EvalM} - K^2\dfrac{\alpha}{\beta}\intz{\heta_M^3\pd{\heta}{k}\EvalM}  
    + K^2\intz{(F-\nu U + D\partial_z^2U)\heta_M\pd{\heta}{k}\EvalM}
\end{aligned}
\end{equation}
With these results, equation (\ref{2D_dKdT2}) governing the slow evolution of $K(T)$ can be solved numerically.
\end{widetext}

\subsection{Prediction of the wavenumber of the marginal mode from differential geometry considerations}

An alternative approach for deriving an evolution equation for the wavenumber of the fastest-growing and simultaneously marginally-stable mode involves the application of differential geometry considerations to express the curvature properties of the $\sigma(T,k)$-landscape (see figure~\ref{fig:sigma_landscape}) along the ridge $\gamma$ (\ref{2D_ridge}). In this section, we show how the propagation of the parabolic point properties along a given direction constrains the shape of the parameterized surface $\Gamma$ along that direction and how an expression equivalent to (\ref{2D_dKdT2}) can be derived solely from the intrinsic properties of the surface itself. 

Considering the parameterized surface shown in figure~\ref{fig:sigma_landscape} that describes the time-varying dispersion relation $\sigma(T,k)$, re-written here for convenience,
\begin{equation}
    (T,k) \rightarrow \Gamma(T,k)=(T,k,\sigma_r(T,k)),
\end{equation}
and its derivatives
\begin{equation}
    \Gamma'_T=(\partial_T\Gamma_1, \partial_T\Gamma_2, \partial_T\Gamma_3)=(1,0,\partial_T\sigma_r),
\end{equation}
\vspace{-0.5 cm}
\begin{equation}
    \Gamma'_k=(\partial_k\Gamma_1, \partial_k\Gamma_2, \partial_k\Gamma_3)=(0,1,\partial_k\sigma_r)
\end{equation}
at any coordinate point $P=(T,k)\in A$, the tangent space $T_p\Gamma$ to the surface $P$ is defined as the vector space spanned by $\Gamma'_T$ and $\Gamma'_k$ evaluated at the point $P$. The tangent space is then uniquely identified by the normal vector $\mathcal{N}$
\begin{equation}
    \mathcal{N} = \dfrac{\Gamma'_T \cross \Gamma'_k}{||\Gamma'_T \cross \Gamma'_k||}
\end{equation}
whose derivatives $\mathcal{N'}_T$ and $\mathcal{N'}_k$ lie in the tangent space $T_p\Gamma$ (since $\mathcal{N}$ is normalized, $\partial_j|\mathcal{N}|^2=2\bra{\partial_j\mathcal{N}}\ket{\mathcal{N}}=0$, for $j=T$ or $k$) and carry information about the curvature properties of $\Gamma$. To obtain a formulation that is invariant under re-parameterization, however, the curvature of a surface is often described using the concept of the \textit{shape operator} rather than via the normal vector $\mathcal{N}$, as we discuss next.

Defining the shape operator, or \textit{Weingarten mapping}, as the endomorphism mapping elements within the tangent space $T_P\Gamma$ onto the directional derivative of the (normalized) surface normal vector $\mathcal{N}$ 
\begin{equation}
     W = W: T_P \Gamma \rightarrow T_P \Gamma\;,
\end{equation}
\begin{equation}
    W(\Gamma'_T)=-\mathcal{N'}_T \quad \text{and}\quad W(\Gamma'_k)=-\mathcal{N'}_k,
\end{equation}
its matrix representation with respect to the basis $\Gamma'_T$, $\Gamma'_k$ is given by 
\begin{equation}
W = 
\begin{pmatrix}
a & c \\
b & d
\end{pmatrix}.
\end{equation}

Here, $W(\Gamma'_T) = a \Gamma'_T + c\Gamma'_k$ and $W(\Gamma'_k) = b \Gamma'_T + d\Gamma'_k$, and the coefficients are given in terms of first and second partial derivatives of $\sigma_r$ with respect to $k$ and $T$, as shown below.

Recalling the definition of the \textit{first fundamental form} $I_p$ of a parameterization as the linear map that restricts the inner product canonically induced in $\reals^3$ to the tangent space $T_p\Gamma$, 
\begin{equation}
   I_p:T_p\Gamma \times T_p\Gamma \rightarrow \reals,
\end{equation}

\begin{widetext}

\label{2D_1st_fund_form}
\begin{equation}
\begin{aligned}
    I_p(u,w) & =  I_p(e\Gamma'_T+f\Gamma'_k, g\Gamma'_T+h\Gamma'_k)\\
    & = eg\braket{\Gamma'_T} + (eh + fg)\bra{\Gamma'_T}\ket{ \Gamma'_k}+ fh\braket{\Gamma'_k}\\
    & \equiv egE + (eh + fg)F + fhG,
\end{aligned}
\end{equation}
and the definition of the \textit{second fundamental form} $II_p$ as the linear map 
\begin{equation}
\label{2D_2nd_fundam_form}
\begin{aligned}
    II_p(u,w) & = \bra{W(u)}\ket{w} = \bra{W(i\Gamma'_T + l\Gamma'_k)}\ket{m\Gamma'_T+n\Gamma'_k}\\
    & = mi \bra{\Gamma'_T }\ket{W(\Gamma'_T)} + ni \bra{\Gamma'_k}\ket{ W(\Gamma'_T)} + ml \bra{\Gamma'_T}\ket{ W(\Gamma'_k)} + nl\bra{\Gamma'_k}\ket{W(\Gamma'_k)}\\
    & = mi \bra{\Gamma''_{TT} }\ket{\mathcal{N})} + (ni+ml)\bra{\Gamma''_{Tk}}\ket{\mathcal{N}} + nl\bra{\Gamma''_{kk}}\ket{\mathcal{N}}\\
    & \equiv miL + (ni+ml)M + nlN
\end{aligned}
\end{equation}

\end{widetext}
(here expressed with respect to the basis $\Gamma'_T$, $\Gamma'_k$), the explicit form of the shape operator $W$ can be constructed from the matrix representations of the these fundamental forms 
as

\begin{equation}
W = 
\begin{pmatrix}
a & c \\
b & d
\end{pmatrix}=
\begin{pmatrix}
E & F\\
F & G
\end{pmatrix}^{-1}
\begin{pmatrix}
L & M\\
M & N
\end{pmatrix} \text{ .}
\end{equation} 

Defining the eigenvectors $v_i$ of the shape operator $W$ as \textit{principal directions} of the tangent space $T_p\Gamma$, and the corresponding eigenvalues $\lambda_i$ as \textit{principal curvatures} (whose product defines the \textit{Gaussian curvature} of $\Gamma$), an expression for the normal curvature $\kappa_n$ of $\Gamma$ in the direction $v_i$ follows from (\ref{2D_2nd_fundam_form})
\begin{equation}
    \kappa_n = II_p(v_i,v_i) = v_iW(v_i) = \lambda_i,
\end{equation}
which, in fact, corresponds to the normal curvature of all the curves on the surface $\Gamma$,  with tangent vector $v_i$ at $P$. 

When considering a specific point $P=M$ on the ridge $\gamma$, where all the points are parabolic points, the conditions (\ref{parab_conditions}) enforce zero first derivatives of the growth rate, and the tangent plane $T_M\Gamma$ is represented by a horizontal plane spanned by the tangent vectors $\Gamma'_T=(1,0,0)$ and $\Gamma'_k=(0,1,0)$ with normal $\mathcal{N}=(0,0,1)$. Therefore, the shape operator $W$ at $M$ simplifies to 
\begin{equation}
  W = 
\begin{pmatrix}
\partial^2_T\sigma_r & \partial_T\partial_k\sigma_r \\
\partial_T\partial_k\sigma_r & \partial^2_k\sigma_r
\end{pmatrix}.
\end{equation}
Rewriting the tangent vector $\gamma'$ to the ridge $\gamma$ at $M$
\begin{equation}
    \gamma' = \bigg(1, \td{K}{T}, 0\bigg) = \Gamma'_T + \td{K}{T}\Gamma'_k,
\end{equation}
corresponding to $\mu'=(1,dK/dT)$ with respect to the basis of the tangent space $T_M\Gamma$,
the propagation of the parabolic-point properties (\ref{parab_conditions}) along the curve $\gamma$ then requires, from a geometrical point of view, enforcing a zero normal curvature $\kappa_n$ in the direction of tangent:
\begin{equation}
\label{2D_zero_kn}
    W\mu' =
\begin{pmatrix}
\partial^2_T\sigma_r & \partial_T\partial_k\sigma_r \\
\partial_T\partial_k\sigma_r & \partial^2_k\sigma_r
\end{pmatrix}  \bigg(1, \td{K}{T}\bigg) = (0,0).
\end{equation}
Consequently, one of the eigenvalues of the shape operator $W$ vanishes, and so must the Gaussian curvature along the curve $\gamma$. Equivalently, the operator $W$ is singular, and the tangent vector $\mu'$ lies in the nullspace of $W$.

After explicit diagonalization of the operator, enforcing the condition of a zero eigenvalue of $W$ yields the following relation between the second derivatives of $\sigma_r$:
\begin{equation}
\label{2D_zero_eigen}
    \left( \partial^2_k \sigma_r\right) \left( \partial^2_T \sigma_r\right) = \left( \partial_T \partial_k \sigma_r\right)^2 \text{.}
\end{equation}
The constraint of zero normal curvature in the direction $\mu'$ (\ref{2D_zero_kn}) yields the two following conditions:
\begin{equation}
    \begin{aligned}
    \partial_T \partial_k \sigma_r \td{K}{T} + \partial^2_T \sigma_r  = 0,\\
    \partial^2_k \sigma_r \td{K}{T} + \partial_T \partial_k \sigma_r = 0,
    \end{aligned}
\end{equation}
which become redundant upon substituting (\ref{2D_zero_eigen}), yielding a single ordinary differential equation for the evolution of $K(T)$:
\begin{equation}
\label{2D_dKdT_diffgeom}
    \partial^2_k \sigma_r \td{K}{T} + \partial_T \partial_k \sigma_r = 0 \text{.}
\end{equation}

The linear ordinary differential equation derived here from differential geometry considerations appears to differ from the formally nonlinear condition (\ref{2D_dKdT2}) derived by second-order Taylor series expansion along the ridge. However, squaring condition (\ref{2D_dKdT_diffgeom}), dividing by $\partial^2_k \sigma_r$ and using (\ref{2D_zero_eigen}) to replace the squared mixed derivative yields (\ref{2D_dKdT2}) exactly. Consequently, the evolution equations derived by Taylor series expansion at quadratic order and by differential geometry arguments are consistent, with the former being essentially the square of the latter. This equivalence reflects the fact that both approaches capture curvature information, either via including the quadratic terms in the Taylor series expansion or via the shape operator. Nevertheless, there are potential practical advantages of using the more abstract machinery of differential geometry.

\begin{enumerate}
    \item The evolution equation for $K(T)$ is linear. In contrast, in the model problem being studied, it is only after computing the expressions for all second derivatives and simplifying that the $(dK/dT)^2$ term in (\ref{2D_dKdT2}) drops out and a linear form is recovered.
    \item The analytical calculations and numerical evaluations necessary to determine all three second derivatives may vastly differ in complexity.  Computation of $\partial^2_T \sigma_r$  is, in general, significantly more involved than accessing $\partial^2_k \sigma_r$ and the mixed term $\partial_T \partial_k \sigma_r$ due to the need to compute the time-derivative of the fluctuation feedback (induced by $\partial^2_TU)$,  which requires differentiation of the eigenfunction $\hphi$ along the ridge and of the amplitude expression (namely, of $\alpha$ and $\beta$, which also involve $U$ and $\hphi$). The determination of $\partial_T^2\sigma_r$ for a state of marginal stability, already a non-trivial task in the 2D model problem, will be significantly more complicated for systems like strongly stratified shear flows, for which the linear operator is not self-adjoint and the eigenfunctions can not be chosen real. Here, the relation between second derivatives (\ref{2D_zero_eigen}) allows  for the elimination of $\partial^2_T \sigma_r$ and thereby for a substantial simplification of the algorithmic implementation. Alternatively, (\ref{2D_zero_eigen}) can be used to assess the accuracy of numerical implementations. 
\end{enumerate}

\subsection{Evolution equation for the wavenumber of the leading stable mode}

The linear evolution of the fluctuations together with the scale separation inherent in the ($\varepsilon\to 0$) limit system render coupling between the fast and the slow dynamics well-defined only when marginal stability is attained. The fluctuation-induced feedback on the mean field does not converge for positive fluctuation growth rates and is zero for negative ones. The former scenario lies outside the realm of the (asymptotically justified) QL approximation, owing to the lack of scale separation, while the latter corresponds to an instantaneous decay of the fast modes on the slow time scale, leading to uncoupled (and, thus, simplified) dynamics governed by \\

\begin{equation}
\label{2D_IVP_Ustab_QL}
\pd{U}{T} = F(z,T) -\nu U+ D \pdd{U}{z},
\end{equation}
\begin{equation}
\label{2D_EVP_etastab_QL}
\sigma \heta =\left(-1 +k^2U -k^4 +\pdd{}{z}\right)\heta \equiv \Lop\heta \text{ .}
\end{equation}

For the given model system, the growth rate $\sigma$ of initially stable fluctuations will gradually increase owing to their interaction with the externally forced and, hence, growing mean field until the fastest growing mode reaches a state of zero growth rate and the coupled QL dynamics ensues. Although the cumulative effect of the fast modes does not affect the slow dynamics, it is nevertheless crucial to track the evolution of the fast modes in order to detect the mode that first approaches the marginally-stable manifold and becomes slaved to the mean field. In the remainder of this section, we derive an algorithm for the prediction of the wavenumber of the fastest-growing \emph{stable} mode.\\

Referring again to the time-varying dispersion relation as a three-dimensional surface, we indicate with $Q = (T_Q,k_Q)$ a point on the landscape for which $\sigma_Q < 0$  and the maximum condition in $k$, $\partial_k \sigma|_Q = 0$, is satisfied. The prediction of the wavenumber $k$ of the fastest-growing stable mode then only requires the propagation of the maximum property in time. 
Expanding in a Taylor series at first order the derivative of the growth rate $\partial_k\sigma$ around the point $Q$,
\begin{equation}
\begin{split}
    \pd{\sigma}{k}(T_Q + \Delta T, & k_Q + \Delta k)\simeq \\ &\pd{\sigma}{k}\EvalQ + \pdd{\sigma}{k}\EvalQ\Delta k + \pdm{\sigma}{T}{k}\EvalQ \Delta T,
    \end{split}
\end{equation}
and enforcing the maximum condition in $k$, an evolution equation for the wavenumber is obtained:
\begin{equation}
    \label{2D_dkdT_stab}
\dfrac{dk_Q}{dT}\,=\,-\dfrac{\partial_T\partial_k\sigma\eval_Q}{\partial_k^2\sigma\eval_Q}.
\end{equation}

From the geometrical point of view, this case substantially differs from the marginally stable one presented in the previous section. While the maintenance of the marginal stability condition imposes constraints on $\sigma$ and both its first derivatives (hence the need for a second-order Taylor expansion of the growth rate), generating the existence of a flat ridge on the landscape, the propagation of the maximum property only constrains the first derivative (with respect to $k$) of the growth rate, allowing for a ridge with a non-zero Gaussian curvature. 

Nevertheless, as for the previous case, the prediction of the wavenumber requires the determination of the second derivatives of the growth rate. These derivatives can be computed using perturbation analysis and solvability conditions at second order, as detailed in the previous section, or via the following procedure: 
\begin{itemize}
    \item take the first partial derivative of the eigenvalue problem (\ref{2D_EVP_etastab_QL}) with respect to $k$ or $T$;
    \item enforce the solvability condition on the parameter dependent boundary-value problem resulting from the previous step to obtain the first partial derivative of the growth rate (with respect to $k$ or $T$); and 
    \item differentiate the first derivative of the growth rate separately with respect to $T$ or $k$ and evaluate the results at the local maximum point $Q$.    
\end{itemize}
The primary differences in the perturbative approach and the procedure outlined above arise
from the order of differentiation, evaluation, and projection onto the null-space of $\Lopc$ (the Fredholm alternative).

While both procedures explicitly lead to the same result for the second partial derivative of the growth rate with respect to the wavenumber $k$, 
\begin{equation}
    \label{2D_sigma_kk_stab}
    \pdd{\sigma}{k}\EvalQ = -8k_Q^2 + 4k_Q \intz{U \heta_Q \pd{\heta}{k}\EvalQ} \text{ ,}
\end{equation}
they seemingly do not yield the same result for the mixed derivative $\partial_k(\partial_T\sigma)|_Q$.

More specifically, the first methodology, which requires the second-order differentiation of the eigenvalue problem (\ref{2D_EVP_etastab_QL}) in $k$ and $T$ evaluated at the point $Q$ and then the application of the solvability conditions, gives 

\begin{widetext}
\begin{equation}
\label{2D_sigma2_kT_stab}
\begin{aligned}
    \pdm{\sigma}{k}{T}\EvalQ &=2k_Q\intz{\bigg(F-\nu U + D\pdd{U}{z}\bigg)|\heta_Q|} + 2k_Q \intz{(U-2k_Q^2)\heta_Q\pd{\heta}{T}\EvalQ}\\
     &+\,k_Q^2\intz{\bigg(F-\nu U + D\pdd{U}{z}\bigg)\heta_Q\pd{\heta}{k}\EvalQ},
\end{aligned}
\end{equation}
\end{widetext}
while the second approach yields the following results, depending on whether the first derivative is taken with respect to $k$ or $T$: 

\begin{widetext}
\begin{equation}
\label{2D_sigma_kT_stab}
    \dfrac{\partial}{\partial k}\bigg(\pd{\sigma}{T}\bigg)\EvalQ = 2k_Q\intz{\bigg(F-\nu U + D\pdd{U}{z}\bigg)|\heta_Q|} + 2k_Q^2\intz{\bigg(F-\nu U + D\pdd{U}{z}\bigg)\heta_Q\pd{\heta}{k}\EvalQ}
\end{equation}

and 

\begin{equation}
\label{2D_sigma_Tk_stab}
    \dfrac{\partial}{\partial T}\bigg(\pd{\sigma}{k}\bigg)\EvalQ = 2k_Q\intz{\bigg(F-\nu U + D\pdd{U}{z}\bigg)|\heta_Q|} + 4k_Q\intz{U\heta_Q\pd{\heta}{T}\EvalQ} \text{ .}
\end{equation}
\end{widetext}

Although seemingly different, the three expressions (\ref{2D_sigma2_kT_stab})-(\ref{2D_sigma_Tk_stab}) in fact can be proven to be identical, confirming the interchangeability of the two approaches. The analytical proof in the case of a self-adjoint operator is provided in Appendix~\ref{append_A}, and numerical evidence can be obtained by confirming that the following relation among  (integrals of) the derivatives of the eigenfunctions is satisfied:

\begin{multline}
    2\intz{U\heta_Q\pd{\heta}{T}\EvalQ} = \\ k\intz{\bigg(F-\nu U + D\pdd{U}{z}\bigg)\heta_Q\pd{\heta}{k}\EvalQ}\text{ .}
\end{multline}

Substituting the expressions (\ref{2D_sigma_kk_stab}) and (\ref{2D_sigma_kT_stab}) into (\ref{2D_dkdT_stab}), the evolution equation for the wavenumber of the fastest growing mode reads

\begin{widetext}
\begin{equation}
     \dfrac{d k_Q}{d T} = \dfrac{2k_Q\intz{(F-\nu U + D\partial_z^2U)|\heta_Q|} + 2k_Q^2\intz{(F-\nu U + D\partial_z^2 U)\heta_Q\partial_k\heta|_Q}}{-8k_Q^2 + 4k_Q \intz{U \heta_Q \partial_k\heta|_Q}}\text{ .}
     \label{2D_dkdT_stab_final}
\end{equation}
\end{widetext}

\cleardoublepage
\section{Numerical Implementation}

\subsection{Implementation and validation of the $k$-prediction algorithm}
To test the numerical implementation of our new algorithm, we focus on the dynamics obtained when a space-and-time varying external force\footnote{$F(z,T)=2.5+0.5\cos(t)\cos(z)+0.5\sin(0.6t)\cos(2z)$}  $F(z,T)$ drives the mean field, with the diffusive and viscous coefficients chosen to be unity ($D = 1$, $\nu= 1$).
We prescribe an initial condition\footnote{$U(z)=1.9$, $\heta(z)=0$, $F(z)=2.5+0.5\cos(z)$} from which a marginally-stable state can be reached in a relatively short time.

For the given forcing and initial conditions, the maximum fluctuation growth rate, associated with a mode with wavenumnber  $k=0.975$, initially is negative ($\sigma=-0.094$). Consequently, the slow field $U$ is updated without feedback from the fast modes according to (\ref{2D_IVP_Ustab_QL}). 
Since the external forcing $F$ drives the continued growth of the mean field $U$, the fluctuations become increasingly less stable until the marginal stability condition is satisfied by the mode with wavenumber $k=1.0012$ at time $T=0.18$. From this moment, the amplitude $A$ of the marginally stable mode is set according to (\ref{2D_Amplitude}), thereby producing a restoring force on the slow field through the (now non-zero) fluctuation feedback that maintains a maximum growth rate (approximately) equal to zero.
Figures \ref{fig:2D_sigmaT} and \ref{fig:kT_search} show the evolution over (slow) time of the locally maximum growth rate $\sigma$ and of the corresponding wavenumber $K$, obtained from the elementary \textit{k-search} QL algorithm.

In practice, due to the finite size of the time step $dT$, the growth rate approaches a finite value slightly larger than zero. As suggested by \citet{Chini2022}, however, a $dT$-independent result can be obtained correcting the fluctuation amplitude (\ref{2D_Amplitude}) by means of a damping factor $\lambda$ when $\sigma$ is above a certain tolerance $\Tilde{\sigma}=5\cdot10^{-5}$. 
Considering the variation of the growth rate between two consecutive time steps $T_{n+1}$ and $T_n$, and enforcing an exponential decay at a rate $\lambda$, we require
\begin{equation}
    \sigma^{n+1}-\sigma^n = ( \Tilde{\alpha}-\Tilde{\beta}|A|^2)dT = -\dfrac{\sigma^n}{\lambda}
\end{equation}
(with $\Tilde{\alpha}=K^2\alpha$ and $\Tilde{\beta} = 2K^4\beta$). Thus, the corrected amplitude satisfies
\begin{equation}
   |A|^2 = \dfrac{\Tilde{\alpha}}{\Tilde{\beta}} + \dfrac{\sigma}{\lambda\Tilde{\beta} dT}\text{ .}
    \label{2D_A2_dfac}
\end{equation}
We note that some level of care is required in specifying the tolerance $\Tilde{\sigma}$ and damping factor $\lambda$. The combination of too low a threshold $\Tilde{\sigma}$ together with too small a damping factor $\lambda$ can drive the growth rate sufficiently close to zero that slightly negative values can be realized. In such cases, our algorithm would instantaneously set the fluctuation feedback to zero, causing artificial discontinuities in the dynamics.

\onecolumngrid\
\begin{center}
\begin{figure}[ht]
\centering
\includegraphics[width=0.6\textwidth]{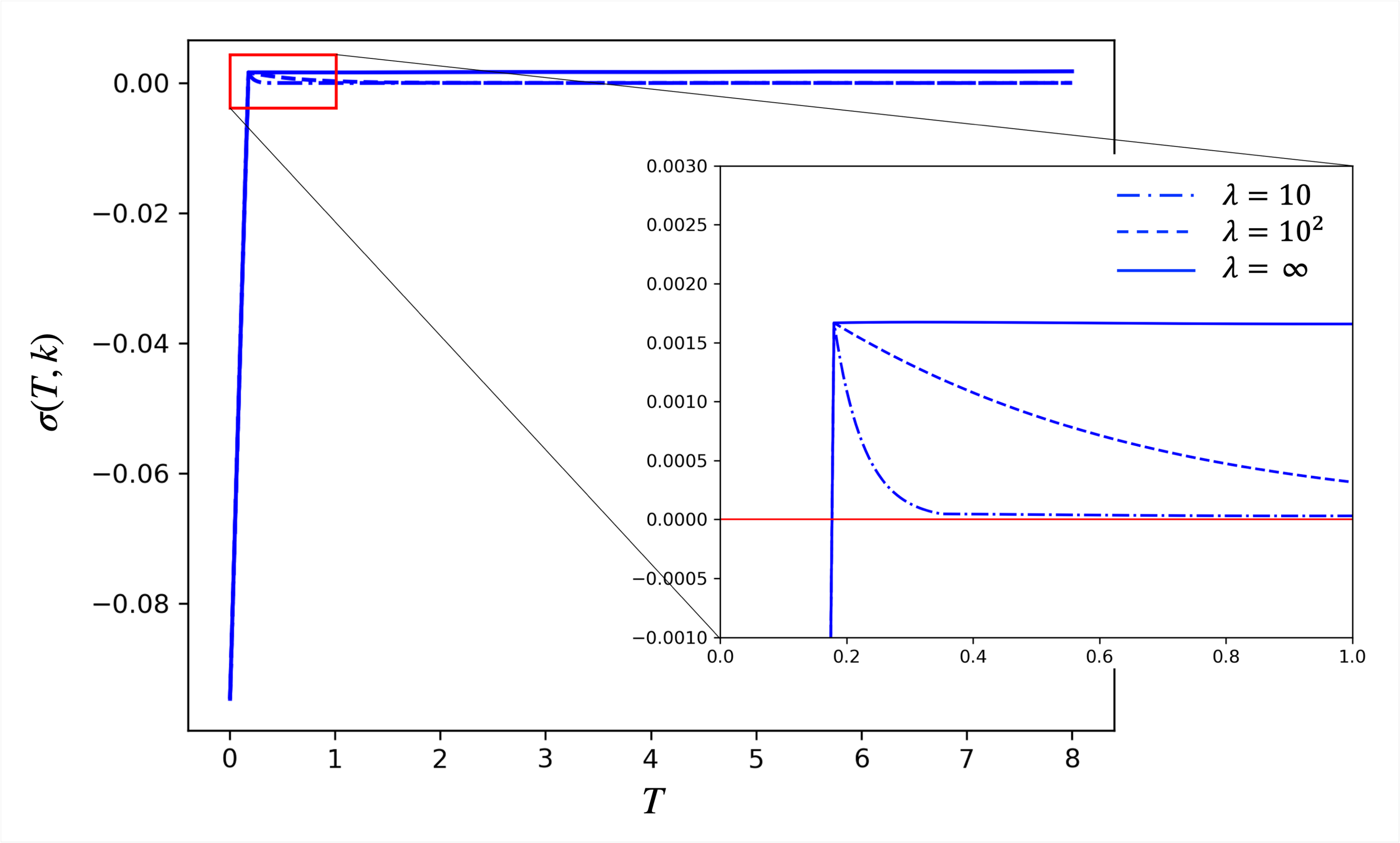}
\caption{Evolution of the growth rate $\sigma$ over slow time $T$. The inset shows the effect of the amplitude correction on the growth rate for two different damping factors $\lambda$ (dashed lines) together with the uncorrected evolution (solid line). }
\label{fig:2D_sigmaT}
\end{figure}

\begin{figure}[ht]
\centering
\includegraphics[width=0.65\textwidth]{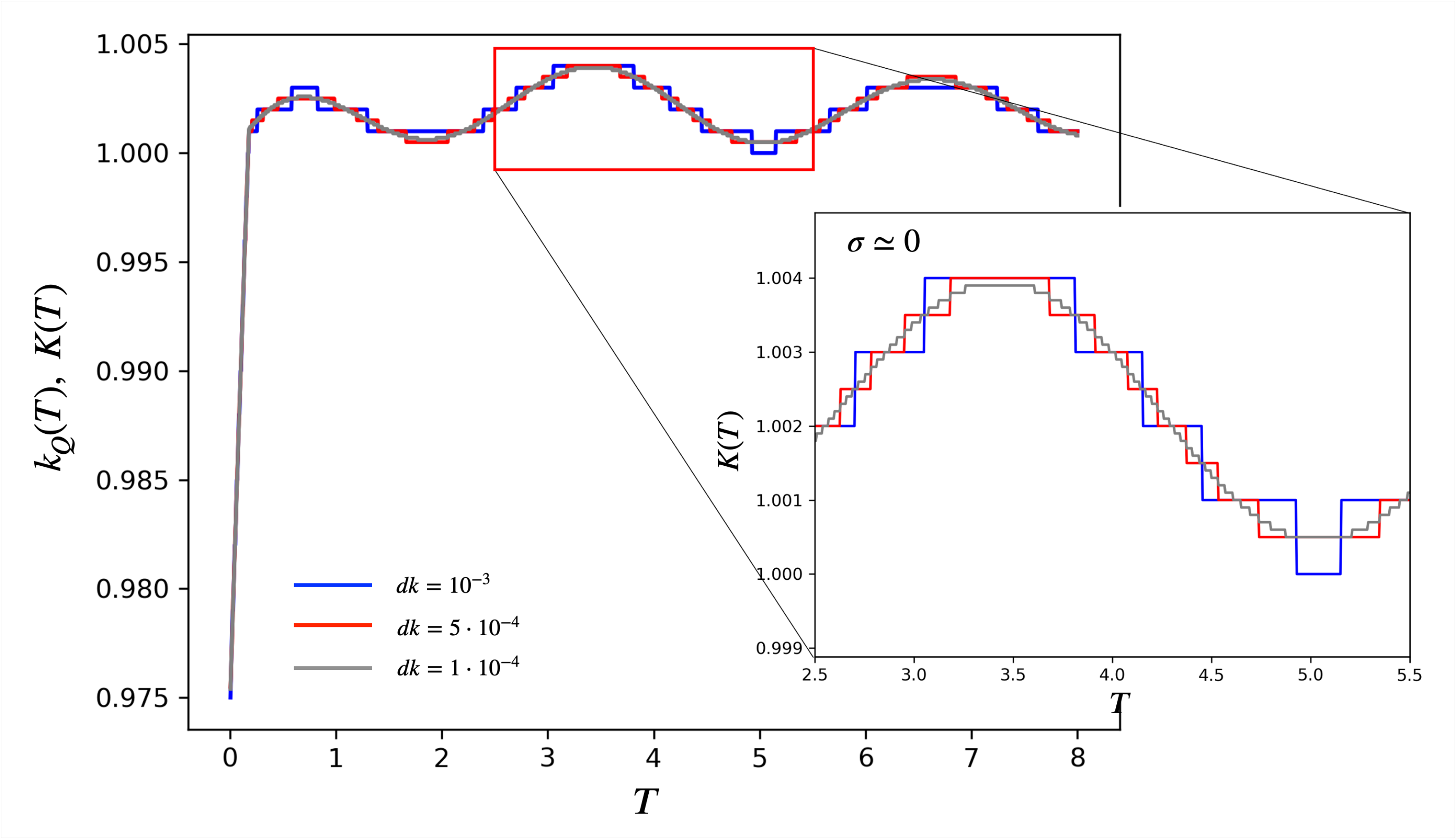}
\caption{Evolution of the wavenumber of the fastest growing mode $k_Q(T)$ (when $\sigma<0$) and of the wavenumber of the marginally-stable mode $K(T)$ obtained from the $k$-$search$ algorithm for various discretizations $dk$.}
\label{fig:kT_search}
\end{figure}

\end{center}
\twocolumngrid

\begin{center}
\begin{figure}[ht]
\centering
\includegraphics[width=0.7\textwidth]{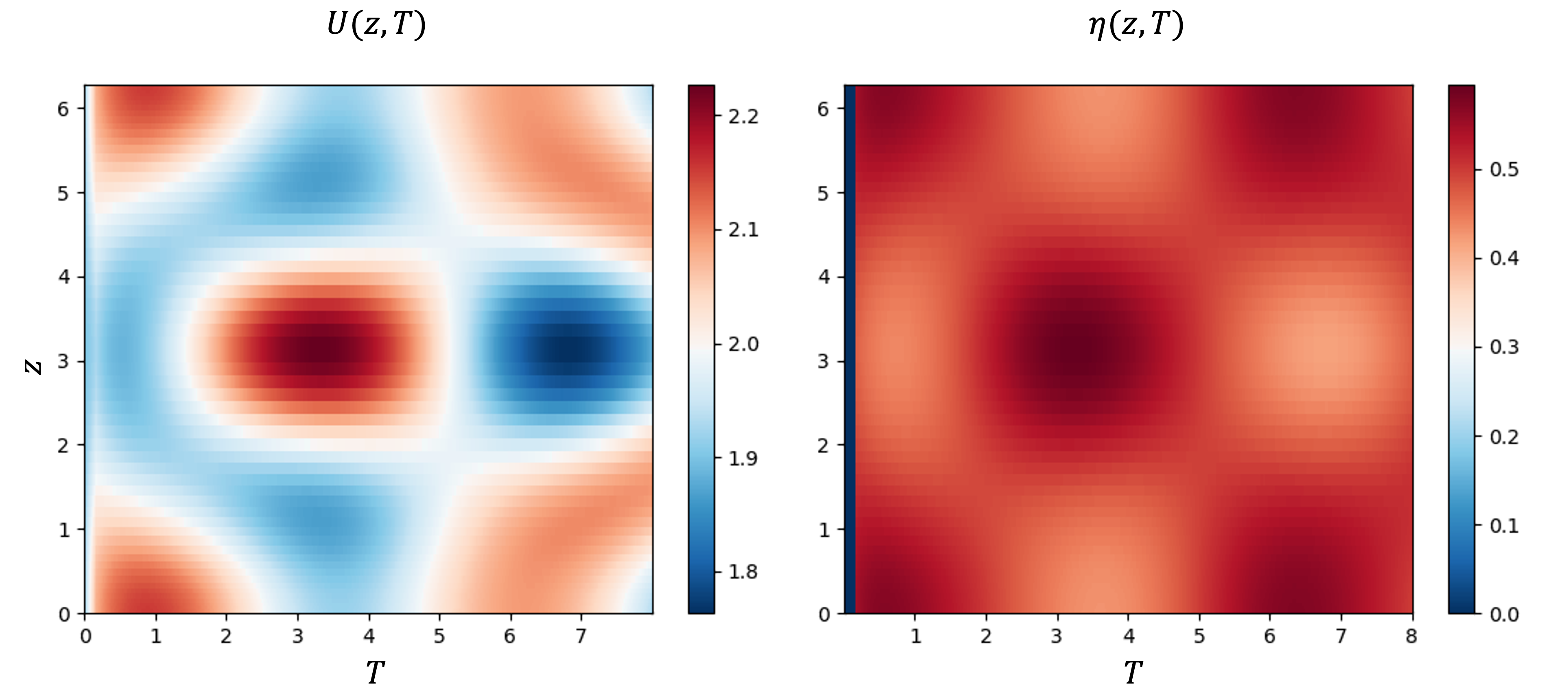}
\caption{Space-time evolution of the mean field $U(z,T)$ (left) and fluctuations $\eta(z,T)=A(T)\heta(z,T)+c.c.$ (right), obtained from the \textit{k-search} QL algorithm.}
\label{fig:2D_U_eta_QLks}
\end{figure}
\end{center}

The marginally-stable state reached by the system is shown in figure~\ref{fig:2D_U_eta_QLks}. Although not an invariant solution of the full system from a strict dynamical systems perspective, this nonlinear state is characterized by a simplified evolution. Specifically, the collapse of the fast modes, now slaved to the marginally stable mean field, gives rise to a low-dimensional dynamics.

\onecolumngrid\
\twocolumngrid\

As evident in figure \ref{fig:kT_search}, the wavenumber of the marginally-stable mode varies in time, oscillating around unity. This variation requires numerous eigenvalue problems, corresponding to different values of $k$, to be solved at each time to identify the peak of the dispersion relation $\sigma(T,k)$.
Moreover, because the QL fluctuation dynamics are autonomous in the streamwise coordinate $\chi$, the search in principle has to be performed using an asymptotically small $k$ increment $dk$ and over a wide range of wavenumbers $k$, dramatically increasing the computational time required for the simulations. This limitation of the \textit{k-search} algorithm becomes of fundamental importance for systems with more complex dispersion relations, where different local maxima may exist and evolve, and for systems where multiple modes can become marginally stable simultaneously, making the development of the \textit{k-prediction} algorithm desirable.

\begin{center}
\begin{figure}[h]
\centering
\includegraphics[width=0.65\textwidth]{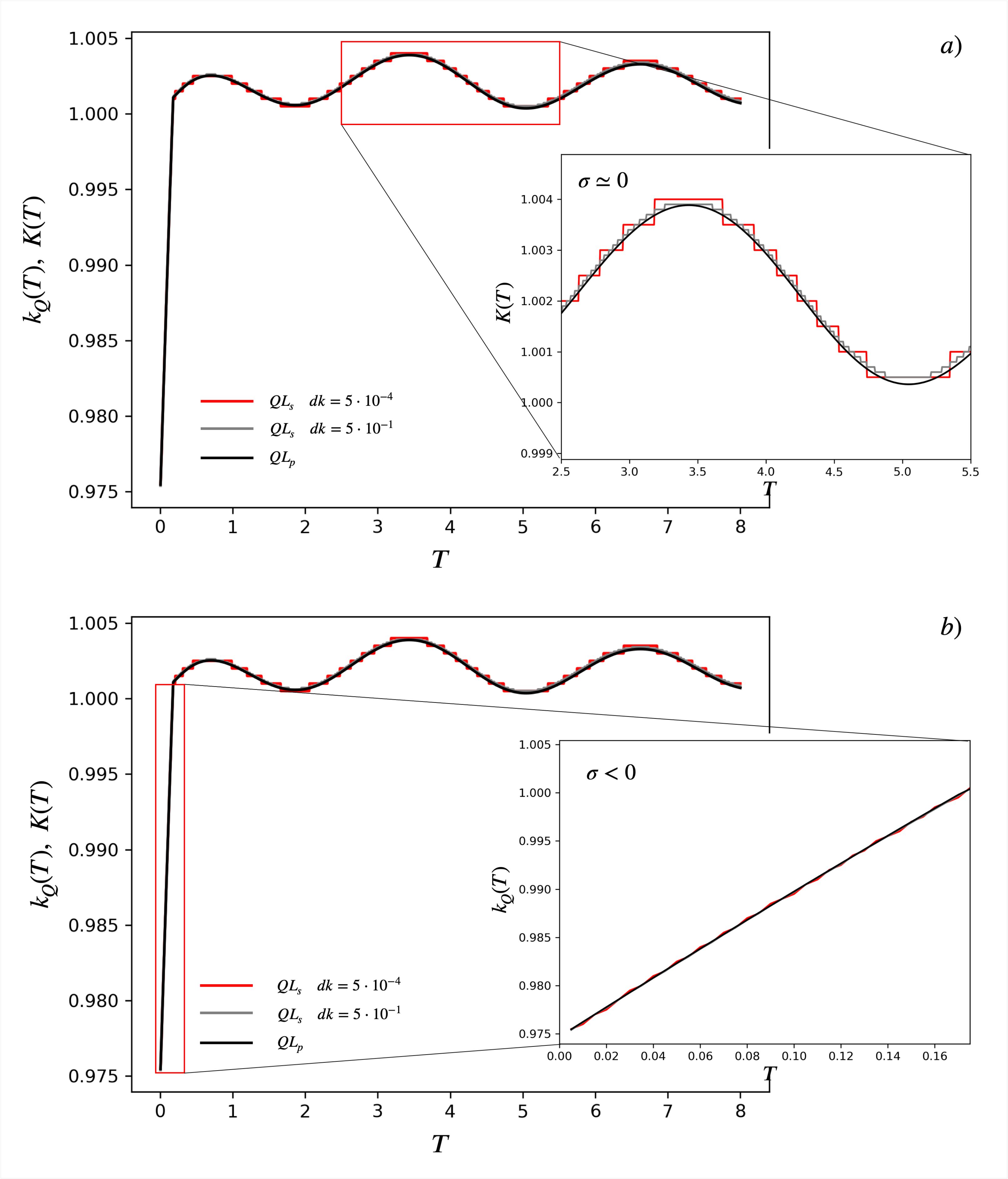}
\caption{Comparison between the evolution of the wavenumber of the marginally stable mode $K$ (top) and of the wavenumber of the fastest-growing mode $k_Q$ (bottom) obtained from the \textit{k-search} algorithm (red and grey curves) and the \textit{k-prediction} algorithm (black curve). The inset in the upper figure shows the comparison in a state of marginal stability, for which the wavenumber is updated according to (\ref{2D_dKdT2}), while the inset in the lower figure shows the comparison for negative growth rates, in which case (\ref{2D_dkdT_stab_final}) is solved. }
\label{fig:kT_pred}
\end{figure}
\end{center}

The efficacy of the new methodology developed in \S~\ref{sec_kpred} for the prediction of the wavenumber of the leading mode both for strictly and marginally stable states ($k_Q$ and $K$ respectively) is shown in figure~\ref{fig:kT_pred}, where output from the two QL algorithms is compared. The results presented below are obtained \emph{without} the use of the amplitude correction (\ref{2D_A2_dfac}), i.e., in the case $\lambda ,\tilde{\sigma}\to \infty$, to highlight the robustness of the \textit{k-prediction} algorithm for non-zero values of the growth rate.
The total number of eigenvalue problems solved during the simulated period is approximately $10^5$ for the \textit{k-search} algorithm (about 60 per time step when $dk=10^{-4}$). During the same period, only $2\cdot 10^3$ eigenvalue problems are solved when the \textit{k-prediction} procedure is employed. In the latter case, only a single eigenvalue problem is solved (i.e., for the maximum wavenumber $K(T)$) at each time step, and only two wide searches over $k$ are performed: the first at time $T=0$ to initialize the wavenumber for the given initial conditions, and the second when the marginally stable manifold is approached, to smoothly connect the two different prediction algorithms (\ref{2D_dkdT_stab_final}) and (\ref{2D_dKdT2}), valid for $\sigma<0$ and $\sigma=0$, respectively.

The evident agreement between the evolution of $k_Q$ and $K$ (figure~\ref{fig:kT_pred}) and the evolution of the mean field obtained from the two different algorithms is even more remarkable considering the modest variations of the wavenumber in this model problem (about $3\%$) and the fact that a forward Euler scheme has been used to update $k$. 

\onecolumngrid\
\
\twocolumngrid\

\subsection{Validation of the QL dynamics against direct numerical simulations}

Finally, we compare the dynamics resulting from the simulation of the fully nonlinear PDE system (\ref{Nonlin_U})--(\ref{Nonlin_eta}), for which the scale separation parameter $\varepsilon=0.02$, to the reduced QL dynamics presented in the previous subsection.
As explained in $\S$~\ref{sec_equations}, the strict QL formulation of the original system is obtained via a streamwise horizontal averaging procedure, yielding an $X$-independent evolution equation for the mean field $U(z)$ and a $\chi$-varying dynamics for the fluctuations $\eta(\chi,z)$, which results in a varying $k$ within the eigenvalue problem formulation. 
For validation and visualization purposes, a fixed-time comparison between the QL simulation and the DNS can be obtained by \emph{a posteriori} choosing the domain size $L_x$ in the DNS of the full system such that $L_x=2 \pi \varepsilon/K(T^*)$, where the wavenumber of the most unstable mode $K(T^*)=1.0033$ at a specific time $T^*=3$ (when the system has already approached the marginal-stability manifold); i.e., $L_x$ is chosen small enough to suppress $x$-variation of the mean field but large enough to accommodate the fluctuation structure.
The 2D fields $U_{QL}(\chi,z,T=T^*)$ and $\eta_{QL}(\chi,z,T=T^*)$ are reconstructed from the QL simulation via 
\begin{equation}
U_{QL}(\chi,z,T^*)=U(z,T^*) 
\end{equation}
and
\begin{equation}
\eta_{QL}(\chi,z,T^*)=A(T^*)e^{(iK(T^*)\chi)} + c.c.
\end{equation}

\begin{center}
\begin{figure}[ht]
\centering
\includegraphics[width=0.6\textwidth]{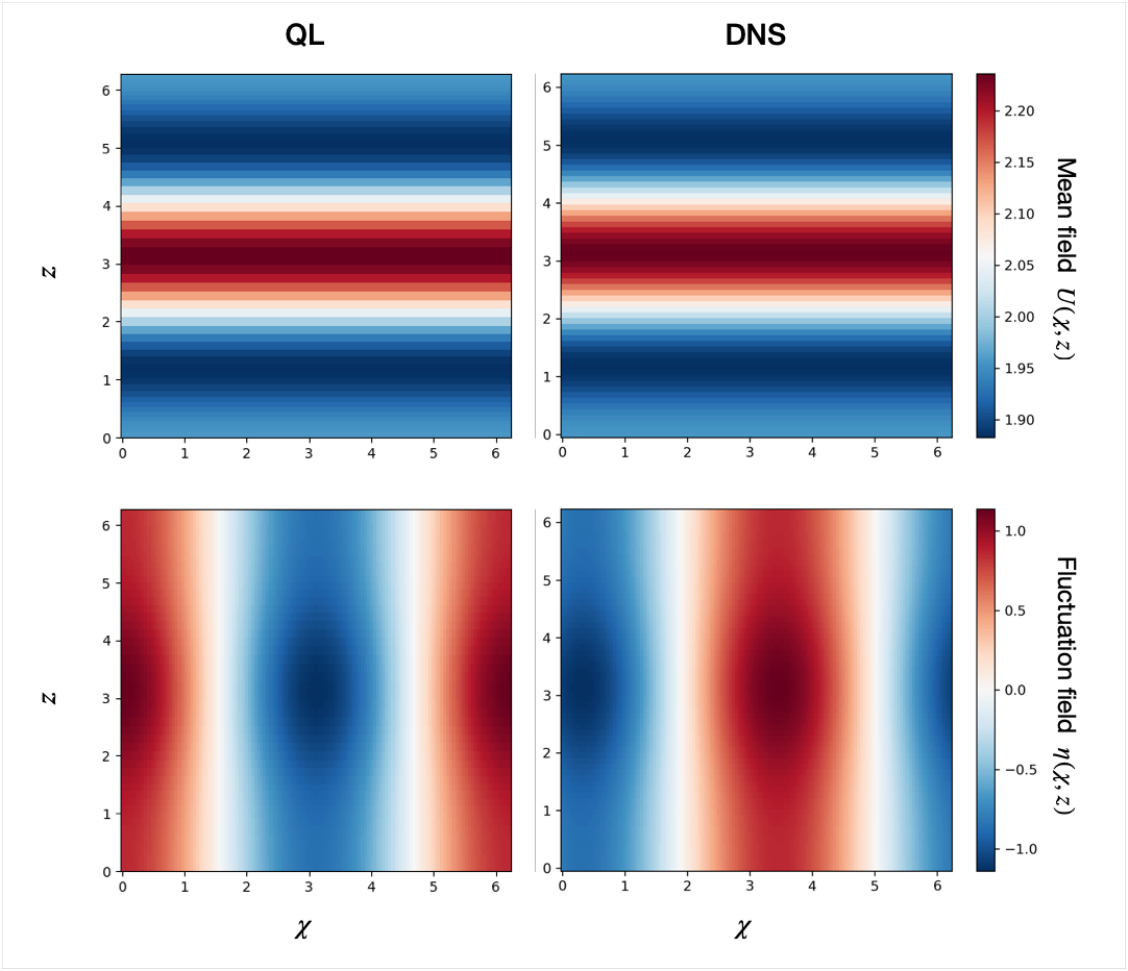}
\caption{Comparison between the mean field $U(\chi,z)$ (top) and the fluctuations $\eta(\chi,z)$ (bottom) obtained from the QL simulation (left) and DNS (right) for fixed time. The QL simulation is performed in a domain  $L_z=2\pi$ with $N_z=64$ modes. The 2D fields are reconstructed at time $T^*=3$ with an $\chi$-independent mean field $U(\chi,z)=U(z)$ and via $\eta(z,\chi,T^*)=A(T^*)\exp(iK(T^*)\chi) + c.c.$ for the fluctuation field (where $K(T^*)=1.0033$). The DNS $z$-domain is identical to that used in the QL simulation, while $L_x=\varepsilon 2\pi/K(T^*)$ in the DNS, with a resolution of $N_x=4$ modes (with a Fourier transform interpolation). }
\label{fig:QL_DNS_comp}
\end{figure}
\end{center}

Despite the finite value of $\varepsilon$ used for the simulation of the full system and the absence of an amplitude correction in the QL simulation (resulting in a `marginal' growth rate $\sigma =0.001$), the comparison given in figure~\ref{fig:QL_DNS_comp} confirms that the reduced QL dynamics faithfully reproduce the behavior of the fully nonlinear PDE system, suggesting the potential for the QL methodology developed here to capture the dynamics of systems that self-tune towards marginally stable states.
The only visible discrepancy, evident in a comparison of the fast modes (figure~\ref{fig:QL_DNS_comp}, bottom row), can be attributed to the translation invariance of the system (\ref{Nonlin_U})--(\ref{Nonlin_eta}) in the $x$-direction that, in conjunction with periodic boundary conditions, allows for an arbitrary $x$-shift in the fluctuation field. Moreover, unlike the results presented in \citet{Chini2022}, where a marginally stable steady state was realized, here the marginally stable state is unsteady, owing to the time-dependent forcing, making the agreement at fixed time arguably more noteworthy.

\onecolumngrid\
\twocolumngrid\

\section{conclusions and outlook}

We have significantly extended an emerging methodology for integrating quasilinear models obtained from asymptotic reductions of forced--dissipative PDEs governing multiscale, spatially anisotropic turbulent dynamical systems including a variety of constrained turbulent fluid flows. Our approach suggests a path towards a dynamical systems description of such multiscale flows, for which classical invariant solutions of the primitive (singularly perturbed) Navier--Stokes equations are not easily identified. The QL reduction of such systems is obtained by decomposing the flow into slow mean and fast fluctuation fields and self-consistently neglecting the subdominant nonlinear fluctuation--fluctuation interactions in the equations governing the evolution of the fluctuations. Within the QL framework, marginally stable manifolds, along which the mean fields evolve such that the dominant fluctuation fields neither grow nor decay on the \emph{fast} time scale, can be identified and followed.

We have investigated a model system that shares certain structural features with the PDEs governing two-dimensional strongly stratified turbulence; specifically, (i)~the fluctuations depend both on a `vertical' ($z$) and `horizontal' ($x$) spatial coordinate; (ii)~the system exhibits dynamics on two time scales that are separated in an appropriate limit; and (iii)~the instability of the slow mean flow occurs on a `fast' horizontal spatial scale that is commensurate with the vertical scale. A multiscale analysis exploiting temporal and spatial scale separation between the mean and fluctuation fields yields a description of the leading-order dynamics comprising an initial value problem for the mean that is coupled to an eigenvalue problem for the fluctuations. Since the mean field is independent of the fast horizontal coordinate, the linear equations for the fluctuations are autonomous in this coordinate. Consequently, a normal mode ansatz is made, which introduces a wavenumber characterizing the (inverse) wavelength of the fluctuations in the horizontal direction. Crucially, only marginally stable eigensolutions can contribute to the feedback of the fluctuations onto the mean. Stable solutions will decay on the fast time scale of the fluctuations and thus not couple to the mean; and unstable solutions will grow without bound and are incompatible with sustained scale-separated dynamics. Consequently, the emergence of coupled dynamics that preserves scale separation requires the fastest growing mode to have an amplitude slaved to the mean so that marginal stability of the mode and thus two-way coupling and scale separation are maintained as the mean evolves under the influence of the fluctuation feedback.  

\begin{figure}[ht]
\centering
\includegraphics[width=0.45\textwidth]{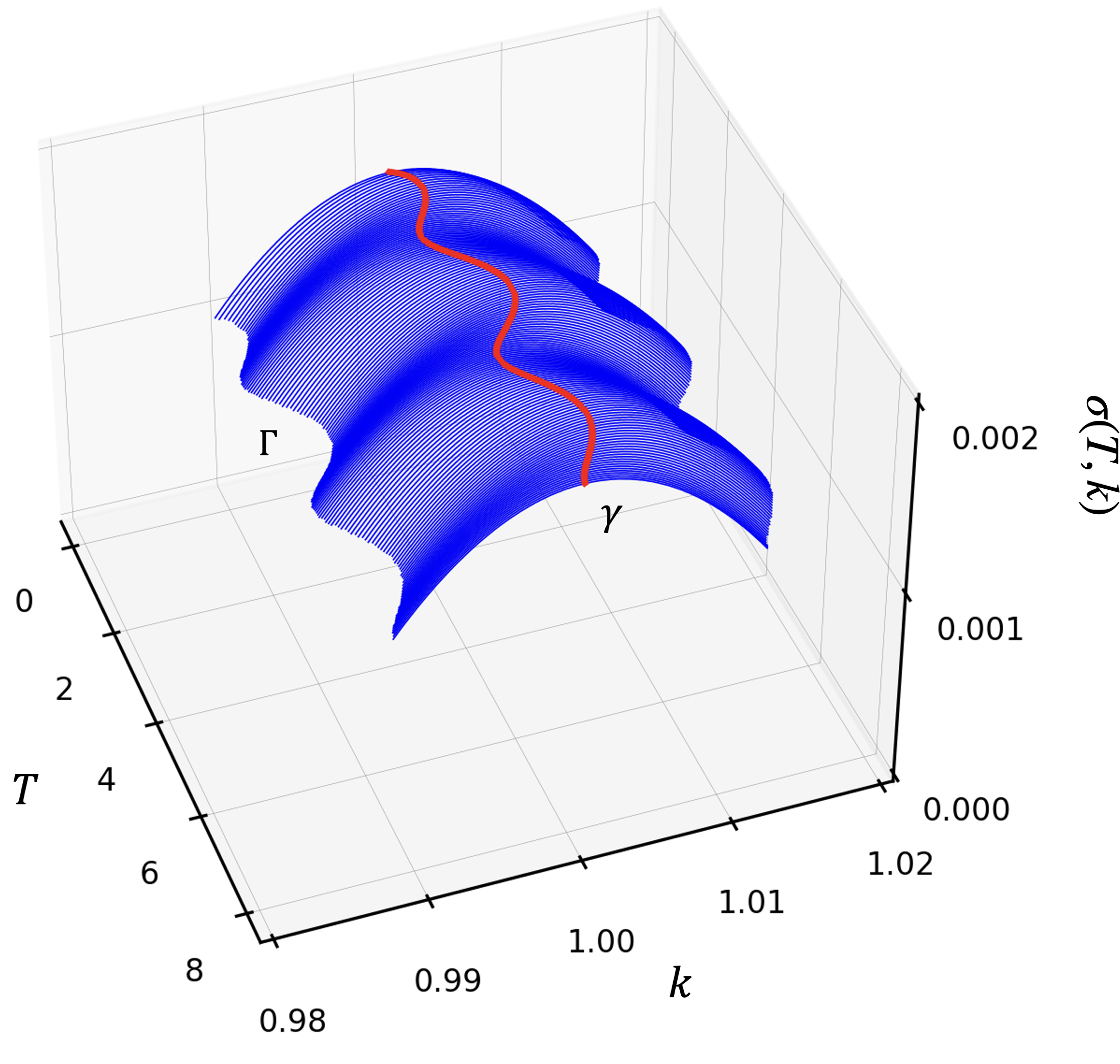}
\caption{Time-dependent dispersion relation $\sigma(T,k)$ obtained from the \textit{k-search} QL algorithm. The blue surface represents the evolution of $\sigma(T,k)$ over slow time while the red curve represents the ridge $\gamma$ given by the evolution in time of the wavenumber of the marginal mode, calculated by (\ref{2D_dKdT2}).} 
\label{fig:2D_sigma3D}
\end{figure}

For the given model system, we demonstrate that the amplitude of the fastest growing mode can be successfully slaved to the mean to preserve marginal stability. Technically, we derive a closed set of differential equations for both the amplitude \emph{and} wavenumber of the evolving marginally stable mode so that marginal stability is propagated in time. Evolving these equations yields a computationally efficient algorithm for following the slowly spatially and temporally adapting, slaved fluctuations. We demonstrate the robustness of our algorithm by comparing the evolution of fluctuation amplitude $A(T)$ and wavenumber $K(T)$ to a method introduced earlier by \citet{Chini2022}. In that work, the fastest growing mode is identified by a computationally expensive brute-force search over an approximately continuous range of $k$, and the amplitude of the fastest growing mode thereby identified is slaved \emph{a posteriori}. As shown in figure \ref{fig:2D_sigma3D}, the proposed algorithm accurately follows the zero-growth-rate ridge in the ($T$, $k$, Re\{$\sigma$\}) landscape, indicating evolution along a marginal stability manifold. Here, a key advantage of the present approach over initial-value simulations of the full primitive (or even full QL) equations is evident; \emph{viz.}, the wavenumber $k$ is not quantized in our formalism, implying that dynamics and pattern formation in spatially extended domains can be accessed.

The marginal stability manifolds define specific state-space structures that guide the dynamics and along which the evolution shows coherence and reduction of complexity. Accordingly, we interpret these marginal stability manifolds as a generalization of the well-studied non-chaotic invariant solutions that have been found to structure state-space in many transitionally turbulent flows. For the state-space of a flow system such as stratified turbulence, decomposing the flow into a mean and fluctuations can be thought of as changing the state-space coordinates so that the two parts of the field live in different dimensions, with some state-space directions being associated with the mean, while the remaining ones describe the fluctuations. This decomposition, itself, does not modify the notion of an invariant solution, corresponding to a trajectory constrained to a subset of state-space that is invariant under the dynamics (e.g., a zero-dimensional point for an equilibrium or a one-dimensional loop for a periodic orbit). Crucially, however, this trajectory must be invariant within the entire state space, i.e., for the mean and fluctuations alike. 

As emphasized previously, when the flow exhibits scale separation in space and time and the QL reduction is valid asymptotically, the fluctuations must be slaved to the mean for scale separation to be preserved and the fluctuations to be sustained. The dynamics collapses in that the fluctuations lose their independence but become parametrically dependent -- or slaved -- to the evolving mean. In the state-space dimensions associated with the fluctuations, the trajectory becomes a graph over the mean field such that the fluctuations may be viewed as an invariant fixed point that parametrically depends on the evolving mean. Since the mean can continue to evolve on the slow time scale, possibly in a chaotic manner, the composite mean/fluctuation system trajectory clearly need not be invariant. Moreover, there is not, in fact, any guarantee that a physically realizable solution for the fluctuation amplitude (i.e., with $|A|^2>0$) exists (see below). Consequently, the marginal stability manifold may end with the trajectory `falling off' the edge of the manifold; for this reason, too, the marginal stability manifold need not be invariant under the dynamics. Obviously, there may be invariant solutions that preserve scale separation and are located within a marginally stable manifold, such as an equilibrium or periodic orbit of the slaved dynamics. However, the marginally stable manifold itself constitutes a different category of state space structure that guides the dynamics and captures coherent structures without necessitating invariance under the dynamics. 
 
In fact, our slow--fast QL formalism yields a criterion to detect the end of these manifolds, as previously noted by \citet{Michel2019} and \citet{Chini2022}: namely, when the parameter $\beta$ becomes negative and no consistent solution for a fluctuation amplitude that preserves marginal stability can be found. Although such $\beta<0$ events are not possible within the model system studied here, for more complex PDE systems these occurrences may indicate fast transient `bursting' events in which the system loses scale separation, undergoes a fully nonlinear transient evolution, and settles onto another marginally stable manifold where scale separation is re-established and the QL description with slaved fluctuations applies. In the context of strongly stratified turbulence, for example, this dynamic corresponds to physically significant, strongly nonlinear mixing events, which are known to be highly spatiotemporally intermittent \cite{Caulfield2021}.

Lastly, we have derived an evolution equation for the wavenumber of any local maximum of the fluctuation growth-rate when this maximum is negative (i.e., `below' marginal stability). Straightforward integration of this equation enables us to follow stable ridges in the temporally evolving growth-rate landscape, and thereby to efficiently identify a mode that may \emph{become} the fastest-growing, marginal mode that subsequently must be slaved to the mean.

Collectively, these developments lay the algorithmic foundation for a dynamical systems description of strongly anisotropic flows admitting scale-separated dynamical evolution. More generally, we believe the concept of marginal stability manifolds should be a useful tool for investigating pattern formation phenomena in spatially-extended domains in systems where patterns are not easily related to classical invariant solutions.

To further develop the methodology and eventually use marginal stability manifolds to describe paradigmatic flows such as strongly stratified 3D turbulence for parameter values characteristic of the ocean or atmosphere, several conceptual challenges first must be addressed. Most importantly, when a manifold ends and the system transiently loses scale separation until another marginal stability manifold is approached, \emph{co-evolution} techniques must be developed that dynamically connect distinct and possibly newly emerging marginal stability manifolds. Although the envisaged phase-space connection is reminiscent of dynamical connections between (unstable) invariant solutions mediated by heteroclinic orbits, here the trajectory exits a marginal stability manifold \emph{not} via an instability \emph{per se} but because the trajectory reaches a boundary where the manifold ceases to exist. For problems where a resolved direct numerical simulation of the original flow equations remains possible, albeit costly, controlled coupling between the QL description for the mean and fluctuations and a DNS of the original equations to be used during the transition between marginal stability manifolds presumably can be achieved. For problems where even short-burst DNS are prohibilitly expensive computationally or otherwise undesirable, alternative methods to co-evolve the mean and fluctuations until marginal stability is recovered must be developed \cite{Ferraro2022}.

In addition, non-smooth transitions between marginally stable modes will occur when the wavenumber of the fastest-growing and therefore slaved mode no longer varies smoothly but when another, \emph{isolated} mode takes over. To handle these transitions or even multiple marginally stable modes contributing to the feedback of fluctuations on the mean, the algorithm for tracking ridges below marginal stability will be particularly useful.

Ultimately, for meaningful application, e.g., to stratified turbulence, the neglected dependence of the mean field on the slow horizontal coordinate $X$ must be reincorporated. This modification, which is roughly tantamount to extending the QL reduction to a GQL formulation \cite{Marston2016}, renders the eigenvalue problem and the associated slaving algorithm dependent on both slow time and the slow horizontal spatial coordinate. Moreover, in three space dimensions, the eigenvalue problem may depend on two fast horizontal coordinates $\chi$ and $\eta$, say, with the prediction algorithm then yielding evolution equations for $k_\chi$, $k_\eta$, and $A$ that parametrically depend on the slow $x$, slow $y$, and slow $t$ variables ($X$, $Y$, and $T$). Nevertheless, the basic architecture of the proposed algorithm will remain unchanged. 

In summary, for singularly-perturbed PDEs admitting slow--fast QL behavior in an appropriate limit, we believe that a dynamical systems description that associates those state-space structures describing significant parts of the flow dynamics with marginal stability manifolds is both feasible and appropriate. The approach carries the hope that marginal stability manifolds together with an algorithm for properly handling the transitions to account for bursting phenomena may yield a reliable description of the full dynamics, and that the methodology will enable the quantitative characterization of flows with strong scale separation at parameter values for which the dynamics cannot be easily evaluated by direct numerical simulation of the associated initial-value problems.\\

\section*{Acknowledgements}

AF and GPC would like to acknowledge the hospitality and financial support of the Woods Hole Summer Program in Geophysical Fluid Dynamics (NSF 1829864), where this work was initiated, along with formative conversations with Prof.~Colm-cille Caulfield (Cambridge University). GPC also would like to acknowledge funding from the U.S. Department of Energy through DE-SC0024572. AF gratefully acknowledges support from the Swedish Research Council (Vetenskapsradet) Grant No. 638-2013- 9243. Nordita is partially supported by Nordforsk.


\appendix*
\appendix
\section{} 
\label{append_A}

Considering a vector space $V:[0,1] \times \reals\rightarrow\reals$ with inner product $\bra{\cdot}\ket{\cdot}: V\times V \to \reals$, we introduce a generic operator $\Lop$ acting on its elements $v$, $\Lop(a,b) : V\to V $ dependent on two parameters $a$ and $b$ with $a,b\in \reals$. This operator is defined self-adjoint, or symmetric, if $\bra{\Lop v_1}\ket{v_2}=\bra{v_1}\ket{\Lop v_2}$, namely if it is identical to its adjoint $L^{\dagger}$. Denoting with $\heta$ the eigenfunction of $\Lop$ (with normalization $\braket{\heta} = 1$) associated with the eigenvalue $\sigma$, the following eigenvalue problem is well defined for any $a,b \in \reals$:
\begin{equation}
\label{EVP_ab}
   \Lop(a,b)\heta(a,b) = \sigma(a,b)\heta(a,b) \text{ .}
\end{equation}

The perturbation of this eigenvalue problem with respect to one or both the parameters $a$ and $b$ allows the derivation of expressions for the corresponding correction to the eigenvalue $\sigma$, via ensuring the solvability of the resulting singular boundary value problem, as detailed in $\S$ \ref{sec_equations}.
When seeking the second-order correction of $\sigma$ with respect to $a$ and $b$, namely $\partial_a(\partial_b\sigma)$, the order of the two differentiations and the imposition of the solvability condition can be exchanged, yielding apparently different expressions.

Denoting the partial derivatives of $\Lop$ and $\sigma$ with the more compact notation $\partial_i(\partial_j \Lop)=\Lop_{ij}$ and $\partial_i(\partial_j \sigma)=\sigma_{ij}$, the three different derivations are summarized below.

\subsection*{ Differentiation w.r.t \texorpdfstring{$a$}{TEXT} and \texorpdfstring{$b$}{TEXT} followed by solvability condition }
Taking the second derivative of (\ref{EVP_ab}) with respect to both parameters $a$ and $b$,

\begin{multline}
    \Lop\heta_{ab} = -\Lop_{ab}\heta -\Lop_a\eta_b -\Lop_b\eta_a \\ + \sigma\heta_{ab} +\sigma_a\eta_b+\sigma_b\eta_a + \sigma_{ab}\heta, 
\end{multline}
and defining the operator $\Lopc = \Lop-\sigma$ (which is singular by construction), yields
\begin{equation}
    \Lopc\heta_{ab} = -\Lop_{ab}\heta -\Lop_a\eta_b -\Lop_b\eta_a +\sigma_a\eta_b+\sigma_b\eta_a + \sigma_{ab}\heta,
\end{equation}
for which the solvability condition (\ref{solvab}) has to be enforced. Taking the inner product with $\heta$,
\begin{equation}
    \label{sigma_ab_1}
    \sigma_{ab} = \bra{\Lop_{ab}\heta}\ket{\heta} + \bra{\Lop_{a}\heta_b}\ket{\heta} + \bra{\Lop_{b}\heta_a}\ket{\heta},
\end{equation}
where $\sigma_a\bra{\eta_b}\ket{\heta}=\sigma_b\bra{\eta_a}\ket{\heta}=0$ due to the preservation of the norm $\bra{\heta_{\phi}}\ket{\heta} = \dfrac{1}{2}\braket{\heta}_{\phi}=0$ (and $\phi = a$ or $b$).

\subsection*{Differentiation w.r.t \texorpdfstring{$a$}{TEXT}, followed by solvability condition and further differentiation w.r.t \texorpdfstring{$b$}{TEXT}}
Taking now the first derivative of (\ref{EVP_ab}) with respect to the parameter $a$,
\begin{equation}
    \label{da_EVP}
    \Lopc\heta_{a} = -\Lop_{a}\heta + \sigma_{a}\heta,
\end{equation}
and projecting the previous expression onto the null-space of $\Lopc$ we obtain
\begin{equation}
    \sigma_{a} = \bra{\Lop_a\heta}\ket{\heta} \text{ .}
\end{equation}

Taking the derivative with respect to the second parameter $b$, a second expression for the mixed derivative of the eigenvalue is obtained:
\begin{equation}
    \sigma_{ab} = \bra{\Lop_{ab}\heta + \Lop_{a}\heta_b}\ket{\heta} + \bra{\Lop_a\heta}\ket{\heta_b},
\end{equation}
which for a self-adjoint operator becomes
\begin{equation}
    \label{sigma_ab_2}
    \sigma_{ab} = \bra{\Lop_{ab}\heta + 2\Lop_{a}\heta_b}\ket{\heta}
\end{equation}
because $\bra{\Lop_a\heta}\ket{\heta_b}=\bra{\Lop_a\heta_b}\ket{\heta}$.

\subsection*{Differentiation w.r.t \texorpdfstring{$b$}{TEXT}, followed by solvability condition and further differentiation w.r.t \texorpdfstring{$a$}{TEXT} }
Repeating the same steps as in the previous case but interchanging the order of the differentiation with respect to $a$ and with respect to $b$ leads to 
\begin{equation}
    \label{db_EVP}
    \Lopc\heta_{b} = -\Lop_{b}\heta + \sigma_{b}\heta
\end{equation}
and 
\begin{equation}
\label{sigma_b}
    \sigma_{b} = \bra{\Lop_b\heta}\ket{\heta} \text{ .}
\end{equation}
after imposing the Fredholm alternative.

Differentiating (\ref{sigma_b}) with respect to $a$, the third expression for the second mixed derivative of $\sigma(a,b)$ reads
\begin{equation}
\begin{aligned}
\label{sigma_ab_2}
     \sigma_{ab}=\bra{\Lop_{ab}\heta + 2\Lop_{b}\heta_a}\ket{\heta}.
\end{aligned}
\end{equation}\text{ .}

The three seemingly different expressions obtained by inverting the order of differentiation and projection onto the null-space of $\Lopc$ can be shown to be identical by demonstrating 
\begin{equation}
\label{to_be_proved}
    \bra{\Lop_{b}\heta_a}\ket{\heta} = \bra{\Lop_{a}\heta_b}\ket{\heta}\text{ .}
\end{equation}

For this purpose, we rewrite (\ref{da_EVP}) and (\ref{db_EVP}) as
\begin{equation}
    0= \Lop\heta_a - \Lop_a\heta +\sigma\heta_a +\sigma_a\heta,
\end{equation}
\vspace{-0.2 cm}
\begin{equation}
    0= \Lop\heta_b - \Lop_b\heta +\sigma\heta_b +\sigma_b\heta,
\end{equation}
and we take the inner product of these expression with $\eta_b$ and $\eta_a$, respectively, obtaining
\begin{equation}
\label{da_EVP_etab}
    0= \bra{\Lop\heta_a}\ket{\heta_b}- \bra{\Lop_a\heta}\ket{\heta_b} +\sigma\bra{\heta_a}\ket{\heta_b} +\sigma_a\bra{\heta}\ket{\heta_b},
\end{equation}
\vspace{-0.2 cm}

\begin{equation}
\label{db_EVP_etaa}
    0= \bra{\Lop\heta_b}\ket{\heta_a}- \bra{\Lop_b\heta}\ket{\heta_a} +\sigma\bra{\heta_b}\ket{\heta_a} +\sigma_b\bra{\heta}\ket{\heta_a}\text{ .}
\end{equation}

Subtracting (\ref{db_EVP_etaa}) from (\ref{da_EVP_etab}) and recalling that $\bra{\heta_a}\ket{\heta_b}=\bra{\heta_b}\ket{\heta_a}$ and $\bra{\heta}\ket{\heta_a}=\bra{\heta}\ket{\heta_b}=0$ because of the normalization yields
\begin{equation}
    \bra{\Lop_a\heta}\ket{\heta_b}- \bra{\Lop_b\heta}\ket{\heta_a} = \bra{\Lop\heta_a}\ket{\heta_b} - \bra{\Lop\heta_b}\ket{\heta_a},
\end{equation}
from which (\ref{to_be_proved}) directly follows upon making use of the self-adjointness of the operator $L$ and the symmetry of the inner product that give $\bra{\Lop\heta_a}\ket{\heta_b}=\bra{\heta_a}\ket{\Lop\eta_b}= \bra{\Lop\heta_b}\ket{\heta_a}$.

\phantomsection

%

\end{document}